\documentclass[fleqn,usenatbib]{mnras}

\usepackage{newtxtext,newtxmath}

\usepackage[T1]{fontenc}

\DeclareRobustCommand{\VAN}[3]{#2}
\let\VANthebibliography\thebibliography
\def\thebibliography{\DeclareRobustCommand{\VAN}[3]{##3}\VANthebibliography}

\usepackage{graphicx}	
\usepackage{amsmath}	
\usepackage{threeparttable}
\usepackage{ulem}
\usepackage{xcolor}

\definecolor{blazeorange}{rgb}{1.0, 0.4, 0.0}
\definecolor{seagreen}{rgb}{0.18, 0.55, 0.34}
\definecolor{rufous}{rgb}{0.66, 0.11, 0.03}
\definecolor{royalfuchsia}{rgb}{0.79, 0.17, 0.57}
\definecolor{scarlet}{rgb}{1.0, 0, 0.0}
\definecolor{royalpurple}{rgb}{0.47, 0.32, 0.66}
\definecolor{darkblue}{rgb}{0, 0, 0.66}
\newcommand\ts[1]{#1}

\title[Alignment]{A Log-Uniform Initial Magnetic Field Distribution Explains Pulsar and Magnetar Populations through Magnetic Inclination Alignment}

\author[T. Shimasue et al.]{
Takumi Shimasue$^{1,2}$\thanks{E-mail: shimasue-takumi@resceu.s.u-tokyo.ac.jp}, Kenta Hotokezaka$^{2}$, Paz Beniamini$^{3,4,5}$
\\
$^{1}$Department of Physics, Graduate School of Science, The University of Tokyo,
7-3-1 Hongo, Bunkyo, Tokyo 113-0033, Japan\\
$^{2}$Research Center for the Early Universe, School of Science, The University of Tokyo,
7-3-1 Hongo, Bunkyo, Tokyo 113-0033, Japan\\
$^{3}$Department of Natural Sciences, The Open University of Israel, P.O. Box 808, Ra’anana 4353701, Israel\\
$^{4}$Astrophysics Research Center of the Open University (ARCO), The Open University of Israel, P.O. Box 808, Raanana 4353701, Israel\\
$^{5}$Department of Physics, The George Washington University, 725 21st Street NW, Washington, DC 20052, USA.\\
}

\date{Accepted XXX. Received YYY; in original form ZZZ}

\pubyear{\the\year{2026}}

\begin{document}
\label{firstpage}
\pagerange{\pageref{firstpage}--\pageref{lastpage}}
\maketitle

\begin{abstract}

The origin of the gap in the observed magnetic field distribution between pulsars and magnetars raises a  challenge to understanding these populations within a unified framework. We analytically show that the gap can be naturally explained by the alignment of the magnetic inclination angle between the magnetic and spin axes. Based on coupled evolution of spin-down and magnetic inclination angle in the plasma-filled magnetosphere, the alignment timescale follows $\tau_\alpha \propto B^{-2}$. Thus, strongly magnetized neutron stars including high-$B$ pulsars and magnetars align more rapidly than pulsars with $10^{12}\,\mathrm{G}$, reducing their beaming fraction and thereby suppressing their observed numbers. However, magnetars are primarily identified through X-ray activity and are therefore relatively less affected by beaming. Taking into account both beaming fraction and luminosity corrections, we reconstruct the initial magnetic field distribution from the observed distribution. We show that pulsars and magnetars do not dictate intrinsically distinct initial distributions, but can instead be understood within a single continuous initial magnetic field distribution, such as a log-uniform distribution.
\end{abstract}


\section{INTRODUCTION}
Isolated neutron stars are divided into several classes based on their primary energy sources, including pulsars, magnetars, \ts{Magnificent seven (M7)} and central compact objects (CCO) \footnote{\ts{The dipole magnetic fields of CCOs are not responsible for their total magnetic field strengths \citep[for a review, see][]{deluca2017}. For example, a crustal toroidal field of the CCO in Kes 79 several orders of magnitude stronger than the dipole field has been proposed to explain its large X-ray pulse fraction \citep{Shabaltas2012}. Moreover, the dipole field itself can be suppressed due to fallback accretion \citep{Ho2011,Vigano2012}.}}. Each class is associated with a characteristic range of \ts{dipole} magnetic field strengths, and these differences are thought to give rise to the diversity in the nature of isolated neutron stars. 
The diversity in the magnetic field strengths and spin periods of newborn neutron stars is suggested to play a crucial role in powering a wide range of high energy transient phenomena, including stripped-envelope supernovae \citep{Nakar2024Natur},
superluminous supernovae \citep[SLSNe;][]{Kasen2010}, gamma-ray bursts \citep[GRBs;][]{Usov1992,Zhang2001ApJ,Thompson2004,Metzger2011MNRAS,Beniamini2017}, fast radio bursts \citep[FRBs;][]{Bochenek2020,Petroff2022,Beniamini2025role,Beniamini2025can}, and ultra long period magnetars \citep[ULPMs;][]{Beniamini2023,Cooper2024,Men2025}.

Observationally, it has long been  recognized that the typical dipole magnetic field strength of Galactic neutron stars is $\sim 10^{12}\,{\rm G}$, as inferred from pulsar spin-down measurements, while magnetars cluster in the range of $\sim 10^{14}$ --$10^{15}\,{\rm G}$ and are far less common than pulsars.  However, the birth rate of magnetars is not necessarily low, given their short lifetimes. In fact,
 \citet{Beniamini2019} found that the magnetar birth rate ($\sim 5\,\mathrm{kyr^{-1}\,galaxy^{-1}}$) is comparable to that of radio pulsars \citep[$\sim 5$ -- $30\,\mathrm{kyr^{-1}\,galaxy^{-1}}$, e.g.][]{Faucher2006}. 
Therefore, it is plausible that these two classes originate from a common shared magnetic field distribution and are understood within a common evolutionary framework (see, e.g., \citealt{Popov2010,Perna2011} for theoretical works and \citealt{Kaspi2005,Camilo2006,Camilo2007} for their common observational aspects).

This picture, however, faces an intriguing observational challenge: there is a clear ``gap'' between the two populations between $10^{13-14}\,{\rm G}$ in $P$-$\dot{P}$ diagram. 
In this context, \cite{Gullon2015,Igoshev2022,Pardo-Araujo2026} suggest that pulsars and magnetars are intrinsically distinct  populations. 
Nevertheless, even if different stellar evolutionary channels or magnetic field generation mechanisms produce a bimodal distribution \citep[e.g.][]{Ferrario2006}, there is no obvious reason why the division between the two birth channels should occur at the same magnetic field scale as the observational dichotomy between pulsars and magnetars.
A less fine-tuned possibility is that this gap arises from selection effects since these two classes are identified through different discovery channels, i.e., pulsars are primarily discovered through large-scale radio surveys, while magnetars are typically identified through burst-triggered wide-field X-ray monitoring \citep[for a review, see][]{Enoto2019}.



Here, we explore the possibility that pulsars and magnetars are born with a common initial magnetic field distribution, rather than arising from a bimodal distribution. 
The clear gap between these two populations may be explained by the evolution of the magnetic inclination angle between magnetic and spin axes, rather than solely by observational selection effects. 

In this study, we analytically show that the coupled evolution of spin-down and magnetic inclination angle driven by magnetic dipole radiation can account for the observed gap between pulsars and magnetars in the magnetic field distribution. In Section~\ref{sec:observation}, we first summarize the observational clues that motivate our idea and explain why alignment is a natural possibility to consider. 
In Section~\ref{sec:analytical}, we analytically describe the physical mechanism of alignment. In Section~\ref{sec:result}, we examine whether the gap is filled once the observed magnetic field distribution is corrected using the beaming fraction predicted. Finally, in Section~\ref{sec:discussion}, we discuss the validity and implications of the alignment scenario, and give conclusions in Section \ref{sec:conclusion}.

\begin{figure*}
    \centering
    \includegraphics[width=0.95\linewidth]{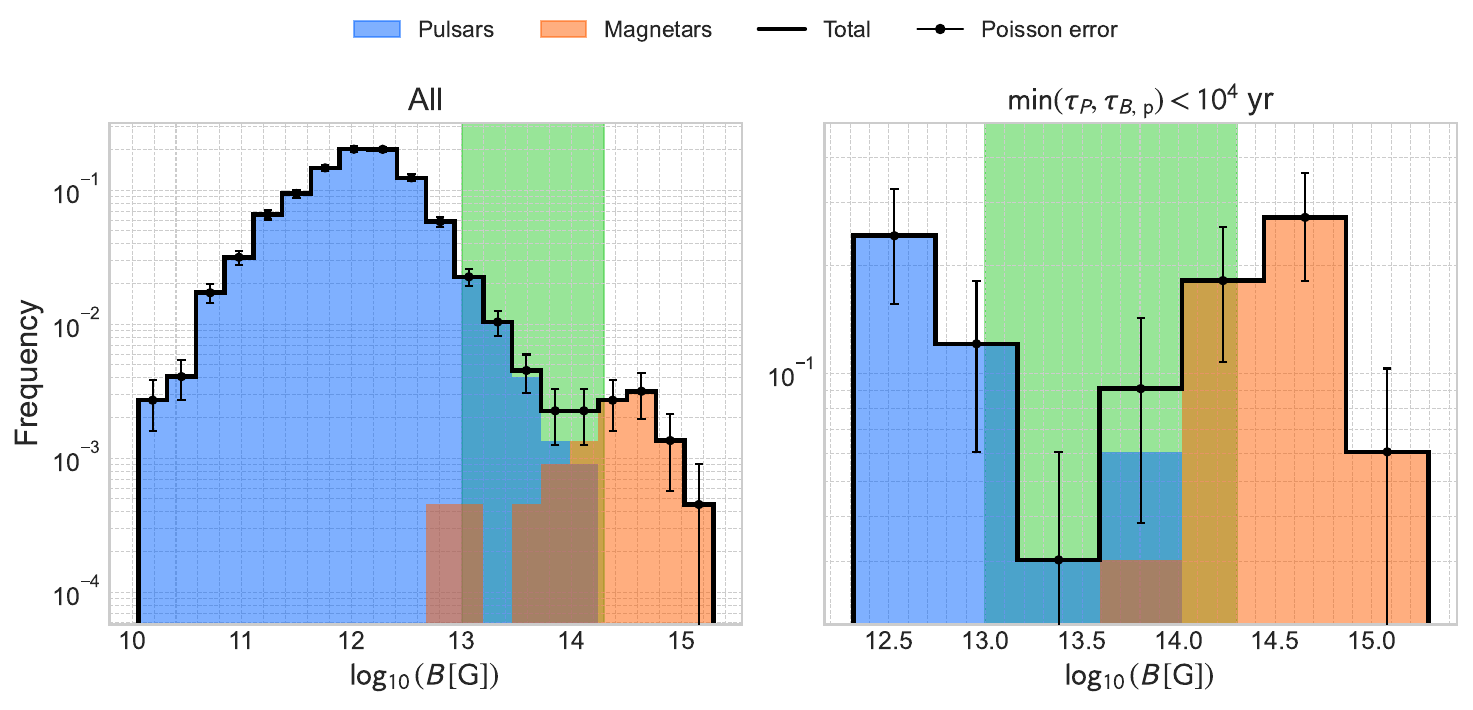}
    \caption{Observed magnetic field distributions of neutron stars.
The panels show the full neutron star sample (left), neutron stars with $\min(\tau_P,\tau_{B,\,\mathrm{p}})<10^4\,\mathrm{yr}$ (right).
Here $\tau_P$ denotes the spin-down age and $\tau_{B,\,\mathrm{p}}$ the magnetic field decay age from persistent emission.
We adopt $\min(\tau_P,\tau_{B,\,\mathrm{p}})$ as a proxy for the true age, following \citet{Beniamini2019}.
Blue and orange histograms represent radio pulsars and magnetars, respectively, while the black solid line shows their sum.
The green shaded region emphasizes the gap between pulsars and magnetars. The error bars show Poisson errors. 
}
    \label{fig:B_hist}
\end{figure*}


\section{OBSERVATIONAL CLUES AND MOTIVATION}
\label{sec:observation}

\begin{table*}
\centering
\caption{Observed magnetic inclination angle (degrees) measurements for magnetars estimated by radio polarization (upper rows) and X-ray measurements, including pulse-profile, polarization, and evolutionary modeling (lower rows).}
\label{tab:inclination}
\scriptsize
\renewcommand{\arraystretch}{0.4}

\resizebox{\textwidth}{!}{%
\begin{tabular}{lccccccccc}
\hline
Name
& $\alpha$
& $\beta^{\rm (a)}$
& $W^{\rm (b)}$
& $\rho^{\rm (c)}$
& $\rho_{\rm psr}^{\rm (d)}$
& $\tau_P^{\rm (e)}$
& $\tau_{B,\,{\rm p}}^{\rm (f)}$
& $\tau_{\alpha}^{\rm (g)}$
& refs.$^{\rm (h)}$ \\
\hline
1E 1547.0-5408$^{\rm (i)}$
& 7--14 & $-5$--0 & 80 & 14 & 3.5 & 0.7--1.4 & 220--460 & 0.04--0.2 & [1][2][3] \\
PSR J1622-4950$^{\rm (i)}$
& 20 & $-10$ & 72 & 19 & 2.4 & 3.5--7.3 & 550--1100 & 2.8 & [4],[5] \\
XTE J1810-197$^{\rm (i)}$
& 8--18 & $-5$--$5$ & 20 & 10 & 2.1 & 8.4--20 & 3000--8000 & 1.6 & [6][7][8][9] \\
Swift J1818.0-1607$^{\rm (i)}$
& $68^{+9}_{-7}$ & $-12.9^{+0.6}_{-0.7}$ & 14 & 28 & 4.3 & 0.24 & 14.5 & 24.7--73.0 & [10],[11] \\
\hline
1RXS J170849.0-400910
& 10 & $30$ & $110$ & $40$ & -- & 9 & 9.3 & 1.2 & [12] \\
SGR J1745-2900$^{\rm (i)}$
& 57 & $0$ & $140$ & 80 & -- & 4.3--9.7 & 2800 & 240 & [13] \\
Swift J1834.9-0846
& 54--70 & -- & $330$ & -- & -- & 4.9 & $>1.4\times10^{4}$ & 189--2526 & [14] \\
1E 2259+586
& $21.2^{+4.6}_{-5.0}$ & $19.6$ & $90$ & 40 & -- & 230 & 1.2 & 84--330 & [15] \\
\hline
\end{tabular}}

\vspace{1.5mm}
\begin{minipage}{\textwidth}
\footnotesize
\setlength{\parindent}{0pt}
${\rm (a)}$ Impact angle between magnetic axis and line-of-sight (degrees).\\[0.4mm]
${\rm (b)}$ Pulse width (degrees).\\[0.4mm]
${\rm (c)}$ Opening angle (degrees) based on the geometry with $\alpha$, $\beta$, and $W$.\\[0.4mm]
${\rm (d)}$ Radio opening angle (degrees) based on Eq.~(\ref{eq:opening angle}).\\[0.4mm]
${\rm (e)}$ Dipole spin-down age (kyr).\\[0.4mm]
${\rm (f)}$ Magnetic field decay age from persistent emission (kyr) derived from \citet{Beniamini2019}.\\[0.4mm]
${\rm (g)}$ Alignment timescale based on Eq.~(\ref{eq:tau_alpha}) (kyr), setting $k_0 = k_1 = k_2 = 1$.\\[0.4mm]
${\rm (h)}$ References: [1] \citet{Camilo2007}; [2] \citet{Camilo2008}; [3] \citet{Lower2026}; [4] \citet{Levin2012}; [5] \citet{Camilo2018}; [6] \citet{Camilo2006}; [7] \citet{Camilo2007XTE}; [8] \citet{Kramer2007}; [9] \citet{Desvignes2024}; [10] \citet{Lower2020}; [11] \citet{Lower2021}; [12] \citet{Zane2023}; [13] \citet{delima2020}; [14] \citet{Li2026}; [15] \citet{Heyl2024}.\\[0.4mm]
${\rm (i)}$ Detected in the radio band, i.e. radio magnetars.
\end{minipage}
\end{table*}

We first summarize the key observational features of the initial magnetic field distributions of pulsars and magnetars, that motivate this study and then outline the proposed alignment scenario.

\begin{figure*}
    \centering
    \includegraphics[width=\linewidth]{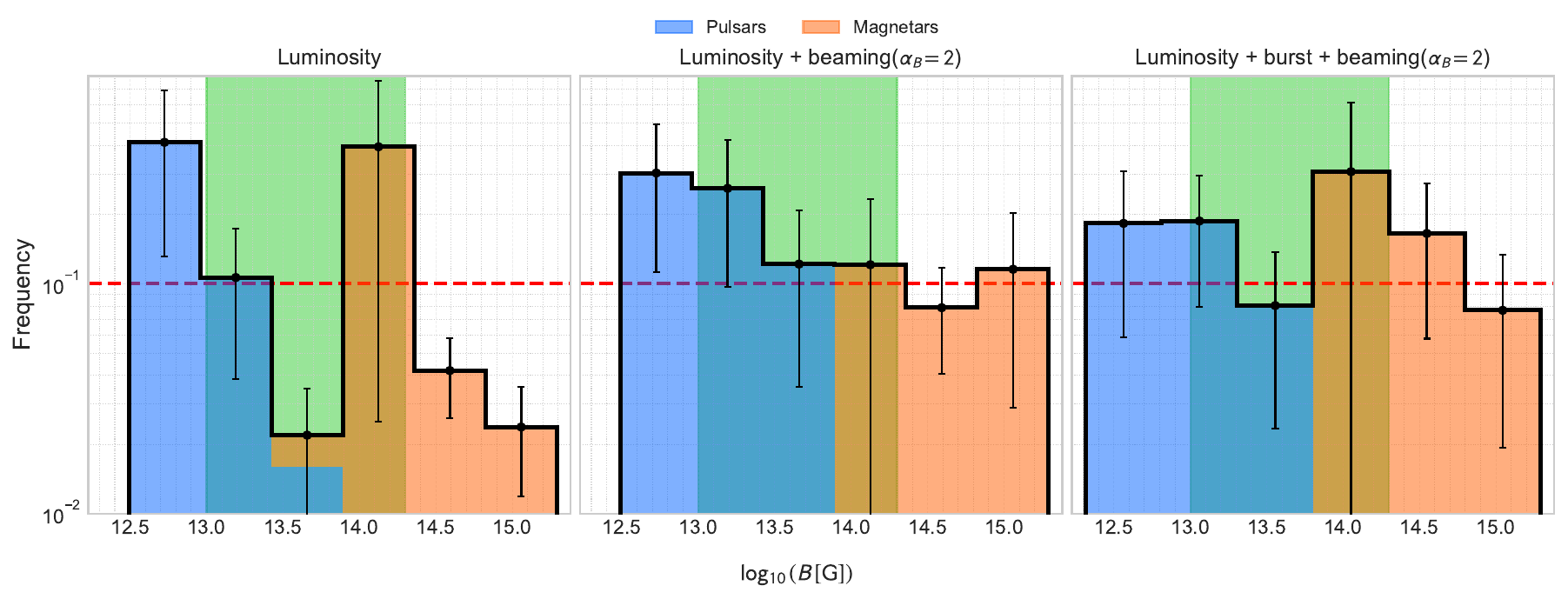}
    \caption{Same as Figure~\ref{fig:B_hist} ({\it{right}}), but weighted by $w_i$ defined in Eq.~(\ref{eq:w_i}), correcting for radio and X-ray survey sensitivities (left), $w_i$ and the beaming correction $f_b\times t$ for $\alpha_B=2$ (middle), Eq.~(\ref{eq:w_i}) for radio pulsars, AXPs and Eq.~(\ref{eq:delta_t_b}) for SGRs with $\eta_{\mathrm{GF}}=1$ and $f_{E,\mathrm{GF}}=0.01$, together with $f_b\times t$ for $\alpha_B=2$ (right).
    }
    \label{fig:B_hist_are_weighted_beaming}
\end{figure*}

\begin{itemize}
    \item \textbf{Birth rate and the gap in the observed dipole magnetic field distribution}: Figure \ref{fig:B_hist} ({\it left}) shows the observed magnetic field distribution of neutron stars. From the sample of all the objects with measured $P$-$\dot{P}$, most neutron stars are observed as pulsars ($\sim 2000$), while magnetars ($\sim 30$) constitute only a small fraction of the total population. That being said, when the sample is restricted to young neutron stars (with ages estimated from the minimum of the spin-down age, magnetic decay time or SNR association when available, following \citealt{Beniamini2019}), the relative fractions of pulsars and magnetars are comparable. \ts{However, there is a clear ``gap'' around $10^{13.5}\,\mathrm{G}$ between pulsars and magnetars, where the fraction reaches a minimum in Figure~\ref{fig:B_hist} ({\it right}).}  \footnote{\ts{We note the lack of a low-$B$ magnetar population among young neutron stars. Low-$B$ magnetars generally have large spin-down ages, are expected to have undergone magnetic field decay. No matter low-$B$ magnetars are present, they are relatively faint and less burst compared with high-$B$ magnetars, as discussed in Sec.~\ref{sec:discussion}. Therefore, they do not contribute significantly in number to the magnetic field distribution restricted to ages of $< 10^4$ yr.}}
    
\item \textbf{Connection between high-$B$ radio pulsars and magnetars}: 
Magnetars are characterized by X-ray activity too energetic to be powered by spin-down losses alone, implying distinct phenomenology from ordinary radio pulsars. 
\ts{The burst activities including short and outburst of low-$B$ magnetars below $B \leq 10^{14}\,\mathrm{G}$ show that the the X-ray activity is not only determined by the surface dipole magnetic field but also by the internal and small scale magnetic field structures of the neutron star \citep[e.g.][]{Tiengo2013,Rodriguez2016,Igoshev2021}.}
However, the balance between spin-down and magnetic decay release changes continuously with initial internal magnetic field. Thus, transitional neutron stars are expected to exist. In fact, several high-$B$ pulsars, most notably PSR J1846$-$0258 and PSR J1119$-$6127, have exhibited magnetar-like activity, including X-ray outbursts, short bursts, and, in some cases, large glitches \citep[for a review, see][]{Kaspi2017}. Moreover, some high-$B$ pulsars show blackbody radiation with temperatures of $kT \sim 0.1$--$0.3\,\mathrm{keV}$, which are higher than that of the typical radio pulsars and similar to those of magnetars \citep{Enoto2019}. Some magnetars are detected 
in the radio band as radio magnetars and share several properties with rotation-powered pulsars, including coherent pulsed radio emission, strong linear polarization, and pulse profiles \citep[for a review, see][]{Enoto2019}.

\item \textbf{Observational selection effects}: The radio luminosity of pulsars is commonly defined as the pseudo-luminosity,
$L_\nu = S_\nu d^2$, 
where $S_\nu$ is the observed mean flux density and $d$ is the pulsar distance \citep{Manchester2005}. Observationally, the pseudo-luminosity shows no strong correlation with the pulsar parameters $P$ and $\dot{P}$. Therefore, luminosity selection alone would not naturally produce a deficit confined to a specific region of the $P$--$\dot{P}$ diagram.

The X-ray persistent luminosity of magnetars is typically 
assumed to be quasi-isotropic,
$L_p = 4\pi d^2 F_X$, where $F_X$ is the flux integrated over the 2-10\,keV band for a given magnetar spectrum \citep{Olausen2014}. Observationally, the X-ray luminosity of magnetars correlate with age. However, even magnetars with ages of $\sim 10^{4\text{--}5}\,\mathrm{yr}$ typically have X-ray luminosities well above $10^{33}\,\mathrm{erg\,s^{-1}}$ \citep{Olausen2014}, making them sufficiently bright to be detected by \textit{Swift}, \textit{Chandra}, and \textit{XMM-Newton}, whose flux sensitivity is $\sim 10^{-14}\,\mathrm{erg\,s^{-1}\,cm^{-2}}$. 
As we describe the details in the Appendix \ref{sec:correction}, the magnetic field distribution reconstructed by applying the luminosity correction backward to the observed distribution shows an even more pronounced gap in Figure \ref{fig:B_hist_are_weighted_beaming} ({\it left}).
Thus, it is difficult to explain the observed gap by luminosity correction solely.

\item \textbf{Magnetic inclination angle}: Observationally, the magnetic inclination angle of radio pulsars can be estimated either through pulse-profile modeling \citep[e.g.][]{Rookyard2015a,Rookyard2015b} or by using the Rotating Vector Model \citep[RVM, ][]{Radhakrishnan1969}. 
Using these methods, the magnetic inclination is measured for several high-$B$ pulsars. PSR J1119-6127, PSR J1513-5908, PSR J1640-4631, PSR J1734-3333, and PSR J1846-0258, respectively,  yields $\alpha = 16$-$21^{\circ}$ \citep[][]{Rookyard2015a}, $13.7^\circ$\citep{Rookyard2015a},   $18.5\pm3^{\circ}$ or  $56\pm4^{\circ}$ \citep[][]{Eksi2016}, $21^\circ$\citep{Nikitina2017} and $10^\circ$ \citep{Li2025}. While, the magnetic inclination angle of typical radio pulsars with $10^{12}\,\mathrm{G}$ is larger than $20^\circ$ \citep{Lyne1985,Tauris1998,Rookyard2015a,Nikitina2017}.
These measurements indicate the magnetic axis tend to align with the spin axis for high-$B$ pulsars. 

The magnetic inclination angles of magnetars can be estimated using radio polarization or X-ray measurements, including pulse-profile modeling \citep{delima2020}, polarization \citep{Zane2023,Heyl2024}, and inference based on an assumed evolutionary model \citep{Li2026}. \ts{In those senses, we note that magnetic inclination angles are subject to observational and model uncertainties for both radio pulsars and magnetars. Especially for magnetars, the number of available estimates is more limited, and several of them rely on model-dependent interpretations of X-ray observations, either from the direction of the magnetic axis or from the geometry of hot spots. Despite these uncertainties and the limited sample, the estimates summarized in Table \ref{tab:inclination} tend to favour relatively small magnetic inclination angles.}
However, Swift J1818.0-1607 is measured to have a relatively large magnetic inclination angle \citep[][]{Lower2020, Lower2021}. This indicates that not all strongly magnetized neutron stars are born with small inclination angles; rather, some objects may begin with large magnetic inclination angle, resulting in a longer alignment timescale as we will see in Sec \ref{sec:analytical}. To check this qualitatively, we compare the reconstructed initial magnetic inclination angle distribution in Figure \ref{fig:alpha0_distribution}, obtained by evolving the observed magnetar population backward based on the method in Sec.~\ref{sec:analytical}. Despite the limited sample size, the reconstructed initial distribution for $P_0 = 1\,\mathrm{s}$ is broadly consistent with an isotropic distribution, with a Kolmogorov--Smirnov (KS) test, $p = 0.52$.
In addition, one of radio magnetars, SGR 1935+2154, which is not included in the table, has been associated with a bright FRB detected in 2020 and several weaker radio bursts in the following years \citep[][]{Bochenek2020,Mereghetti2020, Zhu2023}. Recent studies suggest that active repeaters are aligned magnetar rotators, whereas apparent non-repeaters are misaligned ones, based on arrival-time periodicity, spectro-temporal differences, discovery-rate statistics, energetics, and polarization-angle evolution \citep[][]{Beniamini2025role,Beniamini2025can}.

\end{itemize}

\textbf{Basic idea of this study}:
Motivated by these observational facts, we consider a scenario in which the gap between pulsars and magnetars does not reflect a bimodal initial magnetic field distribution, but instead arises \ts{from the evolution of the magnetic inclination angle.}

In radio pulsars, the emission is beamed around the magnetic axis. As alignment proceeds faster for neutron stars with stronger magnetic fields, the solid angle swept out by the beam becomes smaller, reducing the beaming fraction. As we show in Section~\ref{sec:analytical}, the alignment time scales as $\propto B^{-2}$, implying that strongly magnetized pulsars align more rapidly than pulsars with $10^{12}\,{\rm G}$. Magnetars are primarily identified through X-ray activity powered by magnetic energy dissipation \citep[for a review, see][]{Kaspi2017}, which is expected to be much less affected by beaming than radio pulsar emission. Therefore, magnetic inclination alignment is taken into account, strongly magnetized radio pulsars become progressively invisible, whereas magnetars remain visible. This difference can naturally generate the gap between pulsars and magnetars.

An advantage of this scenario is that it avoids the need for an {\it ad hoc} bimodal initial magnetic field distribution. Instead both pulsars and magnetars can be described within a single continuous distribution, such as a log-uniform or power-law distribution.


\section{Spin and Magnetic Inclination Angle Evolution}
\label{sec:analytical}
The spin-down and magnetic inclination angle evolution of a neutron star may be governed by  \citep{Spitkovsky2006,Philippov2014},
\begin{align}
    \label{eq:Pdot}
    \dot{P} &= \frac{4\pi^2 R^6 B^2}{I c^3 P}
    \left(k_0 + k_1 \sin^2\alpha \right),\\
    \label{eq:alphadot}
    \dot{\alpha} &= -\frac{4\pi^2 R^6 B^2}{I c^3 P^2}
    \left(k_2 \sin\alpha \cos\alpha \right),
\end{align}
where $P$ is the spin period, $\alpha$ is the magnetic inclination angle, $B$ is the dipole surface magnetic field, $I \approx 10^{45}\,\mathrm{g\cdot cm^2}$ is the momentum of inertia, $R = 10^6\,\mathrm{cm}$ is the radius of neutron star.
$k_0$, $k_1$, and $k_2$ are plasma parameters weakly dependent on $R$. 

Combining Eqs.~(\ref{eq:Pdot}) and (\ref{eq:alphadot}),  the equation for the spin-down becomes
\begin{align}
    \label{dotOmegaOmega}
    \frac{\dot{\Omega}}{\Omega}
    = \ts{-}\dot{\alpha}
    \left(
    \frac{k_0 + k_1 \sin^2\alpha}
    {k_2 \sin\alpha \cos\alpha}
    \right).
\end{align}
Integrating Eq.~(\ref{dotOmegaOmega}) yields the following invariant relation \citep[][]{Philippov2014}:
\begin{align}
    \label{Omegak0k1k2}
    \Omega
    \left(
    \frac{\cos^{k_0 + k_1}\alpha}{\sin^{k_0}\alpha}
    \right)^{1/k_2}
    =
    \Omega_0
    \left(
    \frac{\cos^{k_0 + k_1}\alpha_0}{\sin^{k_0}\alpha_0}
    \right)^{1/k_2},
\end{align}
where the subscript ``0'' denotes the initial values.

Using Eq.~(\ref{Omegak0k1k2}), Eq.~(\ref{eq:alphadot}) can be rewritten solely in terms of $\alpha$ as
\begin{align}
    \label{eq:dalpha}
    \dot{\alpha}
    =
    -\frac{1}{\tau_{\alpha}}
    \frac{\sin^{2k_0/k_2 + 1}\alpha}
    {\cos^{2(k_0 + k_1)/k_2 - 1}\alpha},
\end{align}
where the alignment timescale $\tau_\alpha$ is defined as 
\begin{align}
    \label{eq:tau_alpha}\tau_{\alpha}
    &\equiv
    \frac{I c^3P_0^2}
    {4\pi^2k_2 R^6 B^2 }
    \frac{\sin^{2k_0/k_2}\alpha_0}
    {\cos^{2(k_0 + k_1)/k_2}\alpha_0}
    \nonumber\\
    & = \frac{2(k_0 + k_1\sin^2\alpha)\tau_{P} }{k_2} \frac{\sin^{2k_0/k_2}\alpha}
    {\cos^{2(k_0 + k_1)/k_2}\alpha}\nonumber\\
    & =\frac{2(k_0 + k_1\sin^2\alpha_0)\tau_{P,0} }{k_2} \frac{\sin^{2k_0/k_2}\alpha_0}
    {\cos^{2(k_0 + k_1)/k_2}\alpha_0}\nonumber\\
    &\approx
    2 \times 10^3~\mathrm{yr}\,
    B^{-2}_{13}\,P^{2}_{0,-1},
\end{align}
where $\tau_{P,0} \equiv P_0 / 2\dot{P}_0$ is the initial dipole spin-down time. This expression means that the alignment timescale depends only on the initial conditions, $\tau_{P,0}$ and $\alpha_0$.

Here we assume $k_0 = k_1 = k_2 = 1$ motivated by the results of MHD and Force-Free simulations \citep{Philippov2014}. We obtain an analytical solution for Eq. (\ref{eq:dalpha}): 

\begin{align}
\label{eq:alpha_exact}
\ln\!\left(\frac{\sin\alpha}{\sin\alpha_0}\right)
+
\frac{1}{2}
\left(
\frac{1}{\sin^2\alpha}
-
\frac{1}{\sin^2\alpha_0}
\right)
=
\frac{t}{\tau_{\alpha}} .
\end{align}

Where $t$ is the true age.
In the small-angle limit ($\alpha \ll 1$), Eq.~(\ref{eq:dalpha}) is approximated by
\begin{align}
    \label{eq:alpha_t}\alpha(t)
    &\approx
    \left(
    \alpha_0^{-2k_0/k_2}
    +
    \frac{2k_0}{k_2 \tau_\alpha}\, t
    \right)^{-k_2/(2k_0)} \\
    &=
    \alpha_0
    \left(
    1 + \frac{2\alpha_0^2 }{\tau_\alpha}t
    \right)^{-1/2}.
\end{align}

Similarly, Eq. (\ref{eq:Pdot}) becomes 
\begin{align}
    P &\approx P_0 \sqrt{1 + \frac{8\pi^2 R^6B^2k_0}{Ic^3P_0^2}t} \\
    & = P_0\sqrt{1 + B_{13}^2P_{0,-1}^{-2}t_7}.
\end{align}

 For radio pulsars, we assume the polar cap scenario, where the opening angle of pulsars; $\rho$, is described as a function of the present spin period \citep{Sturrock1971,Ruderman1975} and is associated with the bundle of open field lines on the neutron star surface :  
\begin{align}
    \label{eq:opening angle}\rho = k P^{-1/2},
\end{align}
where $k$ is a constant and we set $k = 5^\circ$ based on the result of \citet{Rankin1993, Gould1998}.
We estimate the geometrical beaming fraction \citep{Tauris1998} as a function of the magnetic inclination angle and opening angle :
\begin{align}
    f_b(\alpha,\rho)
&= 
\int_{\max(0,\alpha-\rho)}^{\min(\alpha+\rho,\pi/2)}
d\cos\theta \nonumber\\
& = \begin{cases}
2\sin\alpha\sin\rho,
& \alpha>\rho,\ \alpha+\rho<\dfrac{\pi}{2}, \\[6pt]
\cos(\alpha-\rho),
& \alpha>\rho,\ \alpha+\rho>\dfrac{\pi}{2}, \\[6pt]
1-\cos(\alpha+\rho),
& \alpha<\rho,\ \alpha+\rho<\dfrac{\pi}{2}, \\[6pt]
1,
& \alpha<\rho,\ \alpha+\rho>\dfrac{\pi}{2}.
\end{cases}
\end{align}

For all neutron stars, we apply a magnetic field decay \citep[][]{Colpi2000,Beniamini2019} : 
\begin{align}
    \label{eq:B}B = B_0 \left(1 + \frac{\alpha_B t}{\tau_B} \right)^{-\frac{1}{\alpha_B}},
\end{align}
where $\tau_B\equiv B_{\mathrm{int}}/|\dot{B}_{\mathrm{int}}|$ is the magnetic age and $B_{\mathrm{int}}$ is the internal magnetic field, and $\alpha_B$ is the magnetic field decay index characterizing the dominant field-evolution mechanism. Note that $\alpha_B = 1$  and $2$ correspond to magnetic field decay driven by hall drift and ambipolar diffusion, respectively \citep{Goldreich1992}.
Although there is a region where pulsars and magnetars coexist  on the $P$-$\dot{P}$ diagram, 
here we set the boundary between pulsars and magnetars by equating the spin-down luminosity, $\dot{E}_P = 4\pi^2 I \dot{P}/{P^3} \approx I\Omega^2/2\tau_P$, and the magnetic luminosity, $\dot{E}_{B,\mathrm{int}} = B_{\mathrm{int}}|\dot{B}_{\mathrm{int}}|R^3/6 \approx B_{\mathrm{int}}^2R^3 /6\tau_B$. 
Assuming that the magnetic age is described by a power-law $\tau_B = \tau_{B,14}B_{\mathrm{int},14}^{-\alpha_B}$, a neutron star is classified as a magnetar if it satisfies
\begin{align}
   \label{eq:Pmag} \dot{P}_{\mathrm{mag}} \gtrsim 10^{\frac{2}{\alpha_B}-11} f_\mathrm{int} ^{-\left(\frac{4}{\alpha_B}+2 \right)}
\left(\frac{\tau_{B,14}}{30\,\mathrm{kyr}}\right)^{\frac{2}{\alpha_B}}
P^{-\left(\frac{8}{\alpha_B}+1\right)},
\end{align}
where $\tau_{B,14}$ is normalized at $10^{14}\,\mathrm{G}$, $f_{\mathrm{int}}\equiv B_{\mathrm{int}}/B$ is the scaling factor between the internal magnetic field and the surface dipole magnetic field, set to $f_{\mathrm{int}}=1$ in this study.
We refer to the region satisfying Eq.\,(\ref{eq:Pmag}) as the ``magnetar'' region, while the remaining part of the $P-\dot{P}$ diagram is defined as the ``pulsar'' region.
In the magnetar region, we adopt an effective beaming fraction $f_{b,\,\mathrm{mag}} = 1$ because the magnetars are detected with hard X-ray short bursts which are considered to be isotropic or quasi-isotropic emission.



We assume the initial magnetic inclination angle distribution is isotropic \citep[e.g.][]{Shi2024, Sautron2024}. $\alpha_0 \in [0,\pi/2)$ is generated from $\alpha_0 = \arccos(U)$ with $U \sim \mathcal{U}[0,1]$, corresponding to a uniform distribution in $\cos\alpha_0$.

\begin{figure*}
    \centering
    \includegraphics[width=0.95\linewidth]{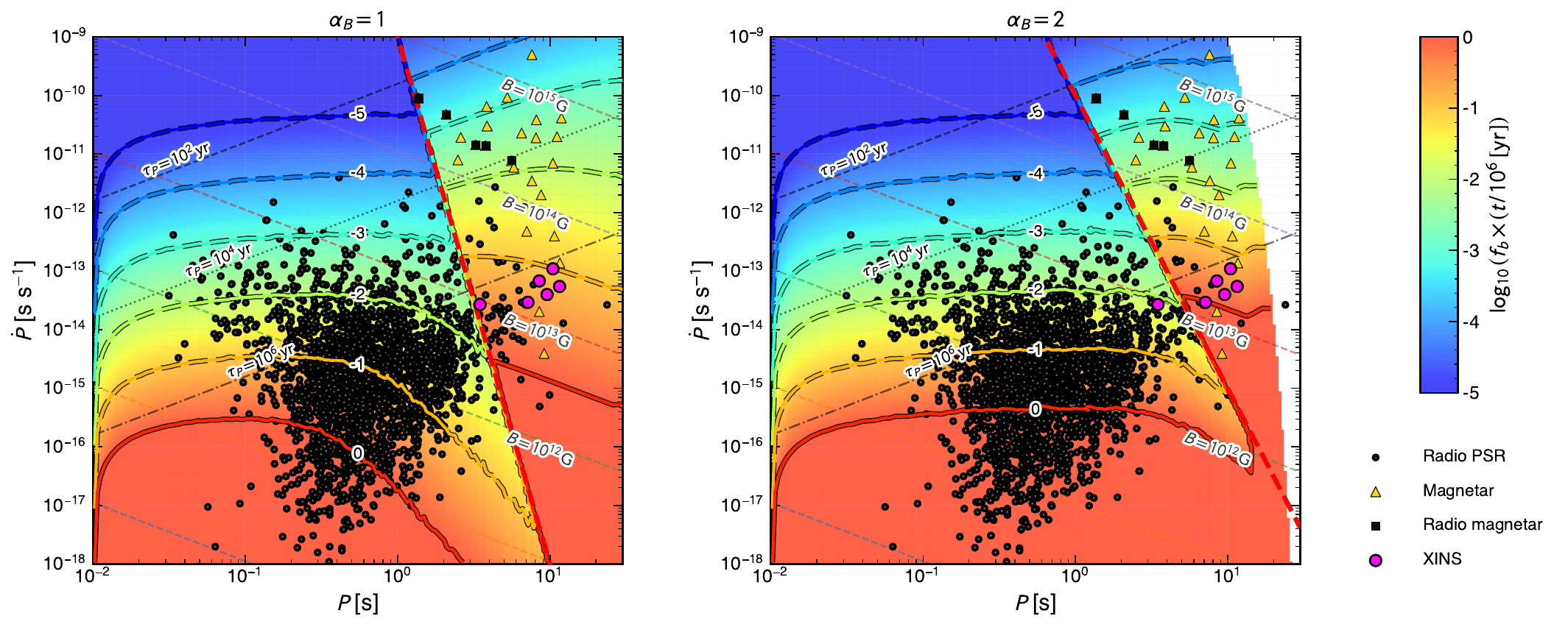}
    \caption{Number ($f_b\times t$) distribution  on the $P$--$\dot{P}$ diagram for two magnetic field decay indices ($\alpha_B = 1$, left; $\alpha_B = 2$, right), assuming $\tau_{B,14} = 30\,\mathrm{kyr}$. The red dashed line shows the boundary between pulsars and magnetars defined in Eq. (\ref{eq:Pmag}). }
    \label{fig:fbt}
\end{figure*}



\begin{figure}
    \centering
    \includegraphics[width=\linewidth]{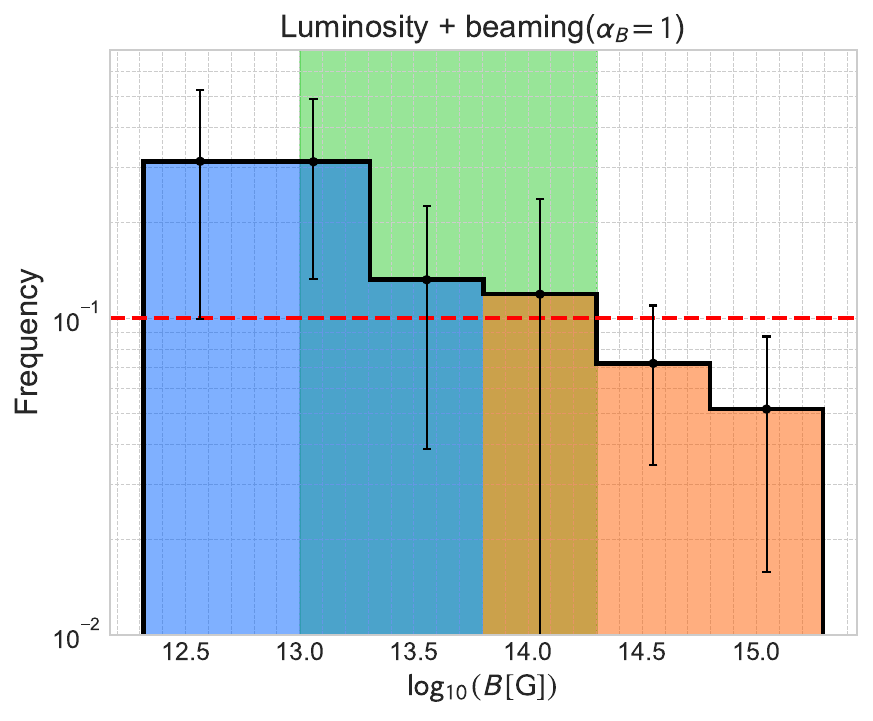}
    \caption{Same as Figure \ref{fig:B_hist_are_weighted_beaming} ({\it{middle}}), but weighted by $w_i$ and $f_b\times t$ assuming $\alpha_B = 1$.}
    \label{fig:B_hist_are_weighted_beaming1}
\end{figure}

\section{Testing the alignment scenario: analytical predictions versus observations}
\label{sec:result}





Here we examine whether the magnetic inclination alignment derived in Sec.~\ref{sec:analytical} can explain the gap by reducing the beaming fraction of strongly magnetized neutron stars and hence decrease their observable number. We then compare the result with the observed magnetic field distribution to test whether the alignment scenario can account for the observed gap. 

Figure \ref{fig:fbt} shows the number distribution generated by the beaming fraction times the true age ($f_b\times t$) on the $P$--$\dot P$ diagram adopting $\tau_{B,14} = 30\,\mathrm{kyr}$ and $P_0 = 0.01\,\mathrm{s}$. Note that the number density on the $P$--$\dot P$ diagram is not determined by the magnetic field itself, instead, it is determined by both the beaming fraction and the true age of neutron stars on the $P-\dot{P}$ diagram. Introducing the boundary between the pulsar and magnetar regions based on Eq. (\ref{eq:Pmag}), we find a transition in the detectability of each neutron star class, which reduces the number density of neutron stars around $10^{13.5}\,\mathrm{G}$ (sky-blue region in Figure \ref{fig:fbt}). In the pulsar region, the observable number density is reduced because of the small beaming fraction. By contrast, in the magnetar region, where beaming effects are negligible, the number density remains relatively high.

We also construct the magnetic field distribution in the same manner as in Appendix \ref{sec:area-weighted}, but now corrected for $f_{\mathrm{b}}\times t$. 
Figures \ref{fig:B_hist_are_weighted_beaming} ({\it{middle}}) and \ref{fig:B_hist_are_weighted_beaming1} show that the magnetic field distribution obtained by applying the luminosity and beaming corrections  and adopting magnetic field decay with $\alpha_B = 2$ and $1$ (see e.g. \citealt{Sautron2025}), are limited to less than $10^4\,\mathrm{yr}$.
Compared with the luminosity corrected distribution ({\it left}) and the luminosity and beaming corrected distribution ({\it middle}) in Figure \ref{fig:B_hist_are_weighted_beaming}, the  gap \ts{around $10^{13-14}\,\mathrm{G}$} disappears. This behaviour is the same for both $\alpha_B = 1$ and $\alpha_B = 2$. 

Importantly, the magnetic field distribution at ages below $10^4\,\mathrm{yr}$ should reflect the initial magnetic field distribution since magnetic field decay has not yet significantly altered the initial field. These results show that an intrinsically bimodal initial magnetic field distribution is unnecessary. A single log-uniform distribution provides a more reasonable description of the initial magnetic field distribution. In addition, magnetic field decay proceeds more slowly for $\alpha_B=2$ than for $\alpha_B=1$. The result for $\alpha_B=2$ favours a log-uniform initial magnetic field distribution while preserving the tail up to $10^{15}\,\mathrm{G}$.

Thus, once the alignment of the magnetic inclination angle is included, the gap in the magnetic field distribution is naturally explained without invoking a bimodal initial magnetic field distribution.

\section{DISCUSSION}
\label{sec:discussion}

\subsection{Validity of the MHD equations and isotropic initial magnetic inclination angle distribution}

We firstly note that the initial distribution of magnetic inclination angles and their evolution remain controversial \citep[for a review, see][]{Li2023}. In the classical vacuum picture, angular momentum is lost through magnetic dipole radiation, spin-down is accompanied by a magnetic torque that drives the spin axis toward alignment with the magnetic axis \citep{Deutsch1955,Ostriker1969,Davis1970,Michel1970}. MHD and Force-Free simulations indicate that pulsar magnetospheres are not a vacuum but are instead filled with plasma \citep{Spitkovsky2006, Philippov2014}. In the plasma-filled case, spin-down occurs even for aligned rotators.
Within these frameworks, the magnetic inclination angle is still expected to decrease with time \citep{Philippov2014}, leading to alignment. The alignment scenario is  broadly consistent with observations \citep[e.g.][]{Tauris1998,Weltevrede2008,Young2010}. 

Nevertheless, it remains unclear whether the alignment scenario can account for all observed properties of the pulsar population.
For instance, some studies suggest that $\alpha$ may evolve toward $\pi/2$ in order to minimise energy release \citep{Beskin1984,Beskin1993} and explain the population of interpulses \citep{Arzamasskiy2017, Kniazev2024}. In addition, the initial magnetic inclination angle may be biased toward low $\alpha$ rather than being isotropically distributed \citep{Rookyard2015a, Rookyard2015b}. However, they do not consider the alignment timescale, which is much smaller than the spin-down age. Observation favouring counter-alignment has also been proposed \citep[][]{Lyne2015}. However, such signatures may simply reflect precession \citep[][]{Zanazzi2015}, and the evolution depends sensitively on the pulse profile and the inclination geometry \citep[][]{Igoshev2020}. 
In this study, we adopt the equations of neutron star evolution describing spin-down and the magnetic inclination alignment based on the plasma-filled magnetosphere \citep{Spitkovsky2006, Philippov2014}.

\subsection{Beaming fraction of radio magnetars}

As we see in Sec \ref{sec:observation}, 
30 magnetars are currently known and as many as 6 radio magnetars have been detected \citep{Olausen2014}, which represents a large fraction even for the aligned neutron stars. If radio magnetars emitted beams in the same manner as canonical pulsars generating close to the magnetic axis, the observed high detection fraction of radio magnetars would be difficult to reconcile with alignment. A natural interpretation is that the radio emission site of magnetars is located at higher latitude than in canonical pulsars, producing a substantially larger opening angle. 
This is observationally supported by position-angle (PA) swings \citep{Dai2019}, circular polarisation \citep{Camilo2007XTE, Camilo2008},  and wide pulse profiles shown in Table \ref{tab:inclination}. A comparison between the radio pulsar opening angle estimated from Eq.~(\ref{eq:opening angle}), and the radio opening angle geometrically calculated from radio magnetar measurements shows that the latter is substantially larger. Recent plasma simulations show that radio emission from magnetars may not originate from open field lines as in canonical pulsars, but instead from closed, twisted magnetic flux tubes anchored near the magnetic poles \citep[][]{Beloborodov2013,Zeng2026}. RVM fitting may not always be valid; nevertheless, many sources still exhibit PA swings consistent with emission from open magnetic flux tubes \citep{Zeng2026} near the polar cap. Therefore, for radio magnetars, the magnetic inclination angle can be reasonably estimated from radio polarisation via RVM fitting, assuming that the emission region is located near the polar cap, as in radio pulsars.

\subsection{Dependence of intial spin period}
\ts{In this study, we adopt the initial spin period as $P_0 = 0.01\,\mathrm{s}$. However, the choice of the initial spin period also affects the observability of an individual
pulsar and the fraction of the neutron star population that evolves into
an aligned state. 
For a nearly orthogonal rotator, $\alpha\simeq90^\circ$, 
$f_{b,\mathrm{misaligned}} = \cos(\alpha-\rho)\approx\sin\rho\approx\rho\propto P^{-1/2}$.
At early times, $P\simeq P_0$, and hence
$f_{b,\mathrm{misaligned}}\propto P_0^{-1/2}$. For a strongly aligned
rotator, $\alpha\rightarrow0^\circ$,
$f_{b,\mathrm{aligned}}
=1 - \cos(\alpha+\rho) 
\approx\rho^2/2
=k^2/2P
\approx k^2/\left(2P_0
\sqrt{1+B_{13}^2P_{0,-2}^{-2}t_5}\right) 
 \propto B^{-1}t^{-1/2}$.
So the leading-order dependence of $f_b$ on $P_0$ vanishes. Although the true age retains a residual dependence on $P_0$, its leading term is $t\simeq\tau_P$. Since $\tau_P$ is also independent of $P_0$ to leading order in this limit, the scaling of $f_b\times t$ remains unchanged. This simple calculation
shows that the beaming correction depends only weakly on $P_0$ in this limit, the shape of the initial magnetic field
distribution is weakly affected by the $P_0$. Nevertheless, the magnetic inclination angle distribution itself remains
sensitive to $P_0$. As we show in Figure~\ref{fig:alpha_distribution}, populations born with shorter initial spin periods contain a larger fraction of aligned pulsars and the observed number of pulsars is reduced.

The typical value of initial spin period is under debate.
Population synthesis allow
initial spin periods ranging between $0.01$--$0.1\,\mathrm{s}$ \citep[e.g.][]{Faucher2006}. The pulsar with supernova remnant (SNR) like the Crab pulsar also provides an
example of a rapidly born neutron star, with a initial period of $20\,\mathrm{ms}$ \citep{Kaspi2002}.
At the point of angular momentum transport from the progenitor of neutron star, the spin of the core before collapse is predicted as slowly rotation \citep{Fuller2019,Ma2019}. However, asymmetric accretion during the supernova can provide angular momentum to the proto neutron star, even for non rotating progenitors \citep{Muller2019,Burrows2023}.

One possible observational probe of estimating the fraction of aligned pulsar might be provided by image-based searches for polarized compact radio sources whose periodic emission has not yet been detected \citep{Sobey2022,Frail2024}. The LOFAR TULIPP survey identified more than
40 known pulsars as polarized LoTSS sources and subsequently searched
30 additional polarized candidates, detecting pulsations from seven
of them and leaving 23 unconfirmed candidates \citep{Sobey2022}.
Similarly, a MeerKAT image-based search selected 30 compelling pulsar
candidates using circular and linear polarization, spectral index,
compactness, variability, and multi-wavelength information
\citep{Frail2024}. Although the unconfirmed fraction cannot be directly interpreted as the
that of aligned one, it suggests that aligned pulsars may constitute
an order-unity fraction of the underlying population, possibly as large
as approximately one half.}

\subsection{Burst correction}

In this study, we focus on young neutron stars ($< 10^4\,\mathrm{yr}$) and show that the observed gap can be explained by magnetic inclination alignment and luminosity correction. However, the distribution limited to $10^5\,\mathrm{yr}$ shows that the fraction of magnetars is lower than that of pulsars (see Figure \ref{fig:mag_dis1e41e5}). Observationally, the magnetar population contains a relatively large number of young, high-$B$ magnetars ($\tau < 10^4\,\mathrm{yr}$, $B > 10^{14}\,\mathrm{G}$; $\sim 20$), whereas the number of old, low-$B$ magnetars is much smaller ($\sim 5$). \citet{Beniamini2019} interpreted this as a consequence of magnetic field decay and showed that, if observational bias is neglected, the observed population favours $\alpha_B=-1$ corresponding to the magnetic decay being proportional to the magnetic field. 
Theoretically, the magnetic decay indices are predicted to be $\alpha_B=0,1,2$ respectively, for Ohmic dissipation, Hall drift, and ambipolar diffusion  \citep{Goldreich1992}. Because the magnetic field decay proceeds more slowly with time for $\alpha_B>0$, a larger number of old magnetars should exist in such cases. This has been pointed out by \citet{Jawor2022}. In the context of their Monte Carlo simulations, they phenomenologically introduced a fade-away function, $S(B) \propto B^{1.73}$, as a function of the dipole magnetic field. They argued that, without the function, $\alpha_B=-1$ is favoured, as in \citet{Beniamini2019}.

Observationally, magnetars are classified into two subclasses, soft
gamma repeaters (SGRs; $\sim15$) and anomalous
X-ray pulsars (AXPs; $\sim10$), and their discovery channels are different \citep[for a review, see][]{Kaspi2017}. SGRs have typically been discovered through burst detections in all-sky surveys with wide-field monitors such as
\textit{INTEGRAL},
\textit{Swift}/BAT, and \textit{Fermi}/GBM. In contrast, AXPs have been identified through follow-up timing observations with {\it Chandra}, {\it XMM-Newton}, and earlier {\it Einstein} observations, often in association with supernova remnants (SNRs) or other counterparts. Notably, among low-B magnetars, two particularly faint sources with weak persistent X-ray luminosities below $10^{30}\,\mathrm{erg/s}$ (\textit{Swift}~J1822.3$-$1606 and SGR~0418+5729) were both discovered with bursts by {\it Swift}/BAT \citep{Cummings2011,Horst2010}. Indeed, both these magnetars are nearby $d\lesssim 2$kpc. This suggests that, at least for old and low-$B$ magnetars, the observational selection is limited by burst detectability and the true number of similar sources in the Milky Way is much larger. The detectability of SGRs is essentially determined by whether at least one burst with energy above the sensitivity occurs during the ongoing duration of the survey. For faint magnetars, the available magnetic energy is smaller, and they become less capable of powering burst activity. In this case, the sources would fail to be detected.

To estimate the burst rate analytically, we define the recurrence time of bursts as 
\begin{align}
    \label{eq:delta_t_b}\Delta t_b \equiv \frac{E_b}{f_{\mathrm{sky}}L_b(E_b)},
\end{align}
where $E_b$ is the energy released in a burst and $L_b(E_b)$ is the time-averaged luminosity associated with the recurrence time of bursts at $E_b$, and $f_{\mathrm{sky}}$ is the sky coverage of the instrument. We assume that all magnetars share the same fluence distribution, which is described by $dN/dE \propto E^{-s}$, where $s$ typically takes $1.6$ \citep[][]{Gogus2000}. Since $E^2dN/dE \propto E^{0.4}$, the total burst energy is dominated by the highest energy burst. Normalizing the burst distribution by the maximum burst energy $E_{b,\max}$ of a given magnetar, Eq. (\ref{eq:delta_t_b}) becomes 

\begin{align}
\Delta t_b
&= \frac{f_{E,\max} f_{\mathrm{dip}} E_{B,\mathrm{int}}}
        {f_{\mathrm{sky}}\eta_{\max} L_p}
   \left(\frac{E_b}{E_{b,\max}}\right)^{s-1}\nonumber \\
&= \frac{\left(f_{E,\max} f_{\mathrm{dip}}\right)^{2-s} E_{B,\mathrm{int}}}
        {f_{\mathrm{sky}}\eta_{\max} L_p}
   \left(\frac{E_b}{E_{B,\mathrm{int}}}\right)^{s-1}.
\end{align}
where $f_{E,\max} \equiv E_{b,\max} / E_{B}$ is the scaling factor between the burst energy and the dipole magnetic energy, $\eta \equiv L_b / L_{p}$ is the scaling factor between the time-averaged burst luminosity and the persistent luminosity, $f_{\mathrm{dip}}\equiv E_{B} / E_{B,\mathrm{int}}$ is the scaling factor between the dipole magnetic energy and total magnetic energy. \ts{As discussed in Sec.~\ref{sec:observation}, the energy available for magnetar bursts depends on the internal magnetic energy reservoir, including both toroidal and poloidal field components, while the energy actually dissipated in MGFs is likely associated primarily with the toroidal component. Therefore, the burst energetics are not uniquely determined by the surface dipole field \citep{Beniamini2025MGF}. We account for this uncertainty through $f_{\mathrm{dip}}$.}
This calculation follows a similar approach to \citet{Beniamini2025MGF}, where the magnetic decay luminosity is derived from a magnetic field decay model characterized by $\tau_B$ and $\alpha_B$ and is converted into burst energy release through the scaling factor $f_{\mathrm{fl}}$. Here, we instead relate the burst energy release to the persistent luminosity, implicitly assuming that the persistent luminosity dominates the magnetic energy output and that the ratio of the time-averaged burst luminosity to the persistent luminosity is independent of age. This leads to a slightly different dependence of the recurrence time of the bursts from that in \citet{Beniamini2025MGF}, for which $\Delta t_b \propto B_{\mathrm{int}}(t)^{2-\alpha_B-2s} \propto L_{B,\mathrm{int}}^{1-\alpha_B/2 - s}$.

For a magnetar located at a distance $d$, the minimum burst energy detectable by an instrument with a fluence sensitivity $\mathcal{F}_{\lim}$ is $E_{b,\min} = 4\pi d^2 \mathcal{F}_{\lim}$. The recurrence time of bursts at the minimum detectable energy can be written as
\begin{align}
    \label{eq:delta_t_b_scaling}\Delta t_b \approx 134\,\mathrm{yr} \, \mathcal{F}_{\lim,-6}^{0.6}  d_{22}^{1.2} L_{p,31}^{-1} \left(\frac{\eta_{\max}}{10}\right)^{-1}\left(\frac{f_{\mathrm{sky}} }{0.1}\right)^{-1}\left(\frac{f_{E,\max}f_{\mathrm{dip}}E_{B,\mathrm{int}}}{10^{43}\,\mathrm{erg}}\right)^{0.4}.
\end{align}
The fluence sensitivities of {\it Swift}/BAT and {\it Fermi}/GBM are $9.8\times10^{-7}$ and $2\times10^{-6}\,\mathrm{erg\,cm^{-2}}$, respectively \citep{Lien2016,Kienlin2020}, and their sky coverages are $17\%$ and $70\%$.
Here we adopt the set-up of \textit{Swift}/BAT  because both old and low-$B$ magnetars, \textit{Swift}~J1822.3$-$1606 and SGR~0418+5729 were detected by \textit{Swift}/BAT. The dipole magnetic field dependence of this expression is $\Delta t_b \propto L_p^{-1}E_B ^{0.4 }\propto L_p^{-1} B^{0.8}$.
With the observed scaling $L_p\propto B^{2.5}$ \citep{Olausen2014}, we obtain
$\Delta t_b\propto B^{-1.7}$, in agreement with \citet{Jawor2022}.

A key issue is how to define the reference scale for the maximum burst energy, ($\eta_{\max}, f_{E,\max}$). Here, we refer to those of giant flares, the most energetic burst events from magnetars. We assume that the statistics of giant flares are a direct extension of the fluence distribution toward the highest-energy end. Table \ref{tab:GF} shows the properties of giant flares observed in the Milky Way. A magnetar giant flare is securely identified when a bright millisecond-long spike is followed by a decaying tail modulated at the spin period of the originating neutron star. Over the past $\sim 50$ yr, three Galactic/LMC magnetar giant flares are confirmed : SGR 1806$-$20, SGR 1900+14, and SGR 0526$-$66. In addition, six extragalactic giant-flare candidates have been identified over the past $\sim 20$ yr, although their classification relies mainly on host-galaxy association rather than on the detection of a pulsating tail \citep{Beniamini2025MGF}. For simplicity, we use only the Galactic/LMC giant flares for fiducial calibration. These events were observed from three of the ten magnetars with $B>5.6\times10^{14}\,\mathrm{G}$, giving a rough recurrence time of
$\Delta t_{\mathrm{GF}}\sim 167\,\mathrm{yr}$ magnetar$^{-1}$.
The time-averaged giant flare luminosity $L_{\mathrm{GF}} \sim E_{\mathrm{GF}}/\Delta t_{\mathrm{GF}}$ of three giant flares are respectively  $L_{\mathrm{GF,1806}}\sim4.6\times 10^{36}\,\mathrm{erg\,s^{-1}}$, $L_{\mathrm{GF,1900}}\sim6\times 10^{34}\,\mathrm{erg\,s^{-1}}$, and $L_{\mathrm{GF,0526}}\sim1.6\times 10^{35}\,\mathrm{erg\,s^{-1}}$.

\begin{figure}
    \centering
    \includegraphics[width=0.9\linewidth]{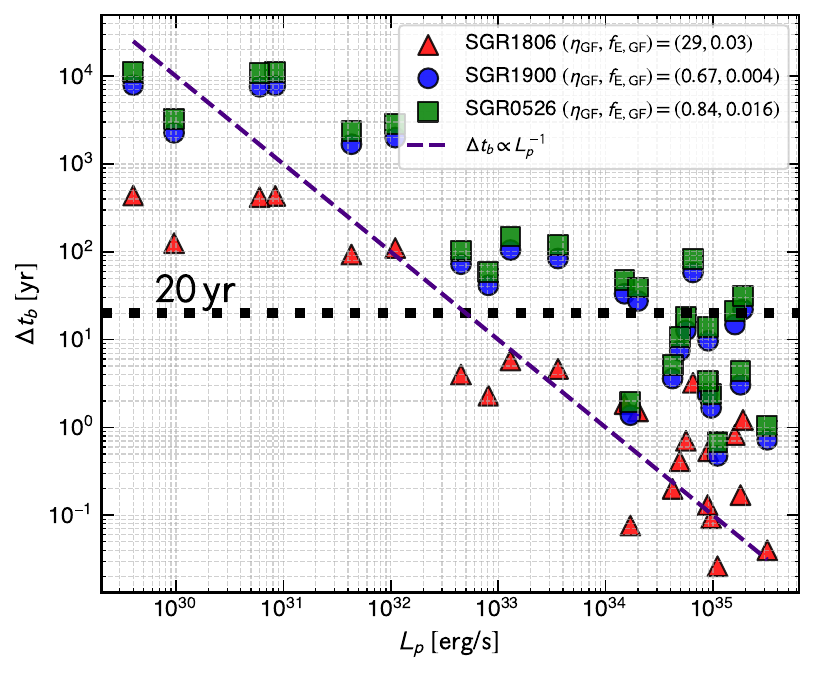}
    \caption{The recurrence time of bursts $\Delta t_b$ derived for all magnetars from Eq.~(\ref{eq:delta_t_b_scaling}) as a function of the persistent luminosity $L_p$. $\Delta t_b$ is adopted with  $(\eta_{\mathrm{GF}},f_{E,\mathrm{GF}})$ from the three observed giant flare magnetars: SGR 1806$-$20 (red triangles), SGR 1900+14 (blue circles), and SGR 0526$-$66 (green squares).}
    \label{fig:delta_t_b}
\end{figure}

Figure \ref{fig:delta_t_b} shows $\Delta t_b$ for all magnetars based on Eq.~(\ref{eq:delta_t_b_scaling}), as a function of $L_p$, using the $(\eta_{\mathrm{\max}},f_{E,\mathrm{\max}}) = (\eta_{\mathrm{GF}},f_{E\mathrm{,GF}})$ of 3 giant flares. The recurrence time of bursts is shorter for young, high-$B$ magnetars, implying more frequent burst activity, whereas it becomes longer for old, low-$B$ magnetars as shown in \citet{Beniamini2025MGF} and mentioned above.
In particular, for the latter population, the recurrence time of bursts is much longer than the observing epoch of \textit{Swift}/BAT. Therefore, a large population of magnetars with low burst activity could exist while still remaining undetected in current burst-based observations. 
This result is also consistent with physically motivated magnetic decay indices of $\alpha_B > 0$. Determining which magnetic decay process is actually favoured by the magnetar population is crucial, but we will study it in future work. 

To verify that the burst correction is particularly crucial for low-$B$ magnetars, we show the magnetic field distribution in Figure \ref{fig:B_hist_are_weighted_beaming} ({\it{right}}) obtained by applying the luminosity correction to AXPs and the burst correction to SGRs on the basis of the magnetar discovery channel classification.
The result is not significantly different from that shown in Figure \ref{fig:B_hist_are_weighted_beaming} ({\it{middle}}), indicating that the burst correction does not strongly affect the inferred distribution of young magnetars. This suggests that young, high-$B$ magnetars retain sufficient magnetic energy to produce at least one burst. We note that the fraction of magnetars around $B\simeq10^{14}\,\mathrm{G}$ is slightly enhanced in Figure \ref{fig:B_hist_are_weighted_beaming} ({\it{right}}). This is mainly due to the some SGRs with low persistent X-ray luminosities despite being young magnetars ({\it Swift }J1834.9$-$0846 and SGR J1745$-$2900). {\it Swift }J1834.9$-$0846 is associated with the SNR W41, whose age is estimated to be $\sim 2\times10^4$ -- $6\times10^5$ yr, longer than $10^4$ yr \citep{Tian2007,nanda2012}. SGR J1745$-$2900 is located close to the Galactic Centre, where the strong interstellar absorption makes it difficult to constrain its persistent X-ray luminosity, leaving only an upper limit \citep{Kennea2013}. These facts indicate that even young, high-$B$ magnetars can appear faint.

\begin{figure*}
    \centering
    \includegraphics[width=\linewidth]{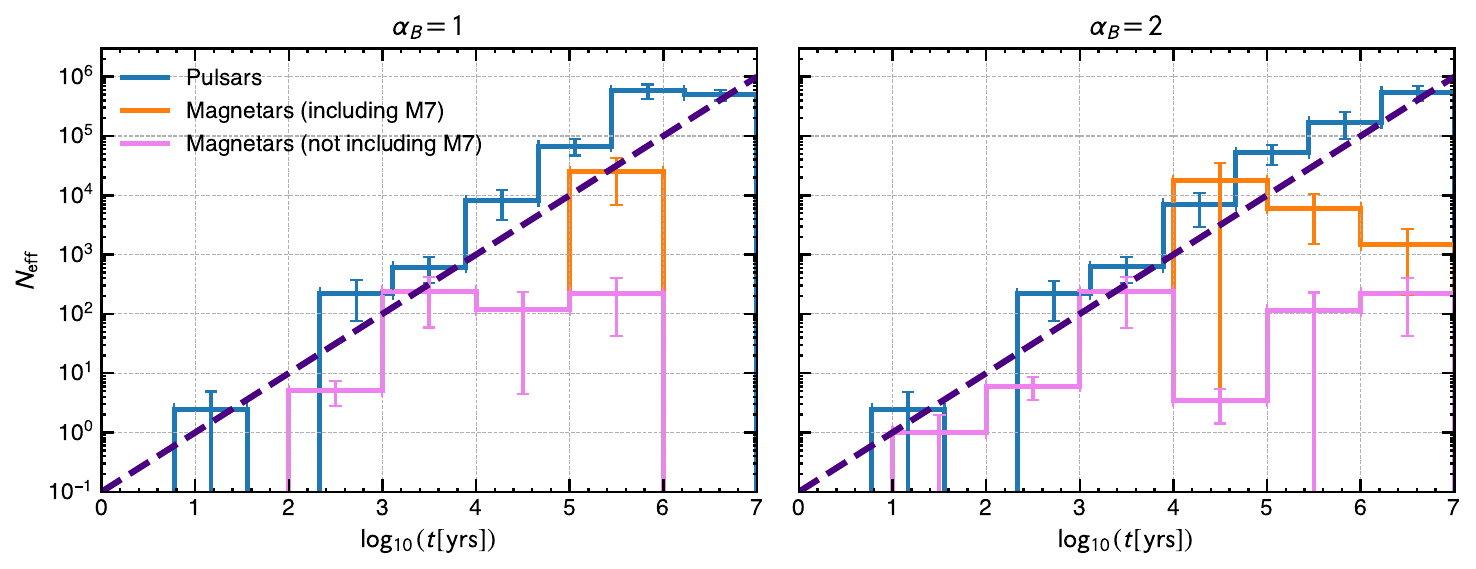}
    \caption{The true age distribution weighted by Eq. (\ref{eq:w_i}) for radio pulsars, AXPs and M7, Eq. (\ref{eq:delta_t_b}) for SGRs with $\eta_{\mathrm{GF}}=1$ and $f_{E,\mathrm{GF}}=0.01$, and $f_b\times t$ assuming $\alpha_B = 1$ (left) and $\alpha_B = 2$ (right).}
    \label{fig:true_age}
\end{figure*}

In addition, based on the method in Sec \ref{sec:analytical}, we also estimate the true ages of the observed pulsars and magnetars and show the reconstructed true age distribution obtained by applying luminosity, burst, and beaming corrections in Figure \ref{fig:true_age}. 
In reconstructing the magnetar distribution, we consider two cases: one including the M7 and the other excluding them \citep[for a review, see][]{Enoto2019}. 
  For the M7, we apply the luminosity correction with the cross section $\sigma_{100\,\mathrm{eV}}$, motivated by the observations; pulsed thermal emission and strong magnetic fields resembling AXPs and high-$B$ pulsars, which are plausibly interpreted as magnetar descendants \citep{Popov2010}. 
Comparing the magnetar cases with and without the M7, we find that the case including the M7 is more consistent with the relation $N_{\mathrm{eff}}\propto t$ up to ages beyond $10\,\mathrm{Myr}$ which indicates a constant birth rate. The effective numbers in the true age distributions of pulsars and magnetars agree within a factor of about two. This suggests that the birth rates of pulsars and magnetars are also comparable within a factor of about two, consistent with previous results from population synthesis \citep[][]{Beniamini2019,Pardo-Araujo2026}. Moreover, the shape of the distribution is insensitive to the choice of the magnetic field decay index. This is because the observed population mainly reflects sources before magnetic field decay significantly occurs, making the effect of magnetic decay less apparent in the reconstructed true age distribution.

\begin{table*}
\centering
\caption{Observed properties of giant flare magnetars and energetics}
\label{tab:GF}

\resizebox{\textwidth}{!}{%
\begin{tabular}{l l l l l c c c c c c c c}
\hline
Name & $E_{\mathrm{GF}}{}^{\mathrm{(a)}}$ &$L_{\mathrm{GF}}{}^{\mathrm{(b)}}$ & $B{}^{\mathrm{(c)}}$& $E_{B}{}^{\mathrm{(d)}}$&$L_{\mathrm{p}}{}^{\mathrm{(e)}}$& $\eta_{\mathrm{GF}}{}^{\mathrm{(f)}}$& $ f_{E,\mathrm{GF}} {}^{\mathrm{(g)}}$ & 
$\tau_P{}^{\mathrm{(h)}}$ & $\tau_{B,\mathrm{p}}{}^{\mathrm{(i)}}$ & $\tau_{B,\mathrm{GF}}{}^{\mathrm{(j)}}$

&$d{}^{\mathrm{(k)}}$& Reference${}^{\mathrm{(l)}}$ \\
\hline
SGR 1806-20  & $2.3\times 10^{46}$ & $4.6\times 10^{36}$ & $2\times 10^{15}$ & $7\times 10^{47}$  & $1.6 \times 10^{35}$ & 29 & 0.03 & $0.16-1.45$ & $6.1-55$ & $0.6-5.3$&8.7 & [1]\\
SGR 1900+14 & $3\times 10^{44}$ & $6\times 10^{34}$ & $7\times 10^{14}$ & $8\times10^{46}$ & $9\times 10^{34}$ & 0.67 & 0.004 & $0.46-1.3$& $8.8-25$& $30-83$& 12.5 & [2]\\
SGR 0526-66 & $8\times 10^{44}$ & $1.6\times10^{35}$ & $5.6\times 10^{14}$ &$5\times10^{46}$ & $1.9\times 10^{35}$ & 0.84& 0.016 & $1.9-3.4$& $10-17$& $26-44$&53 & [3],[4]\\

\hline
\end{tabular}%
}

\vspace{2pt}
\begin{minipage}{0.95\textwidth}
\footnotesize
${\mathrm{(a)}}$ Total isotropic flare energy (erg). \\
${\mathrm{(b)}}$ Time-averaged flare luminosity associated with the recurrence time of giant flare ($\mathrm{erg \,s^{-1}}$). \\
${\mathrm{(c)}}$ Dipole magnetic field strength (G).\\
${\mathrm{(d)}}$ Dipole magnetic energy  (erg). \\
${\mathrm{(e)}}$ X-ray persistent luminosity $2-10\,\mathrm{keV}$ ($\mathrm{erg \,s^{-1}}$). \\
${\mathrm{(f)}}$ $\eta_{\mathrm{GF}} \equiv L_{\mathrm{GF}} / L_{p}$, the scaling factor between the persistent luminosity and the time-averaged flare luminosity. \\
${\mathrm{(g)}}$ $f_{B,\mathrm{GF}} \equiv B_{\mathrm{GF}} / E_{B,\mathrm{dip}}$, the scaling factor between the dipole magnetic energy and the flare energy. \\
${\mathrm{(h)}}$ Dipole spin-down age (kyr). \\
${\mathrm{(i)}}$ Magnetic field decay age from persistent emission (kyr) derived from \citet{Beniamini2019}. \\
${\mathrm{(j)}}$ Magnetic field decay age from giant flare emission (kyr) derived from \citet{Beniamini2019}. \\
${\mathrm{(k)}}$ Magnetar distance (kpc). \\
${\mathrm{(h)}}$ Reference : [1] \citet{Palmer2005}; [2] \citet{Feroci2001}; [3] \citet{Mazets1979}; [4] \citet{Kaplan2001}\\

\end{minipage}
\end{table*}

\subsection{Prediction for high-$B$ pulsars}

If the alignment scenario is correct, several observational predictions for high-$B$ pulsars follow;

\begin{itemize}
    \item \textbf{Small polarization angle swing}: high-$B$ pulsars are expected to show small and shallow polarization angle (PA) swings \citep[e.g.][]{Beniamini2025role}. It can be tested by examining the rotation phase pattern of the PA. Since the RVM directly connects the PA swing to the magnetic inclination angle, this provides an important diagnostic of the alignment scenario.
    \item \textbf{Large distance}: The numbers of high-$B$ pulsars should be reduced due to their small beaming fraction. Therefore, high-$B$ pulsars could show a different distance distribution from pulsars with $B\sim10^{12}\,{\rm G}$, with a relative preference for larger distances owing to the larger surveyed volume.
    \item \textbf{PWN associations}:The alignment scenario also predicts that some PWNe should lack detectable pulsed emission from their central pulsars, since approximately isotropic PWN emission can reveal pulsars whose beams do not intersect with line-of-sight \citep[e.g.][]{Shimasue2026}. Indeed, HESS J1554-550 \citep[][]{Eagle2022} and 2HWC J1953+294 \citep{Coerver2019} have been identified as PWNe based on the presence of X-ray nebulae and SNRs, even though their central pulsars have not been directly detected. Their ages are estimated from the Sedov-Taylor solution for evolved SNRs to be $18\,{\rm kyr}$ and $20\,{\rm kyr}$, respectively. An additional illustrative example is PSR~J1640--4631, associated with HESS~J1640--465, whose dipole magnetic field is $B=1.4\times10^{13}\,{\rm G}$ and spin-down age is $3.35\,\mathrm{kyr}$ \citep{Gotthelf2014}. Although the central pulsar has been detected, pulsations have been observed only in X-rays, with no radio pulsations detected. High-energy emission, especially non-thermal X-ray and gamma-ray emission, is generally produced in a broader magnetospheric region such as outer-gap, making it observable over a wider range of viewing angles. Thus, even if the pulsar is aligned, the non-thermal X-ray pulsation is observed. 
    The alignment scenario therefore predicts that many PWNe without detected central pulsars should host high-$B$ pulsars.
\end{itemize}

\section{CONCLUSION}
\label{sec:conclusion}
We have shown that magnetic inclination alignment offers a natural explanation for the observed gap between pulsars and magnetars in the magnetic field distribution. The centrepiece is that the alignment timescale decreases with magnetic field strength, $\tau_\alpha \propto B^{-2}$, so that strongly magnetized radio pulsars become rapidly aligned and hence increasingly difficult to detect because of their reduced beaming fraction. By contrast, magnetars remain observable through their X-ray isotropic bursts.


These results therefore support a common evolutionary framework in which pulsars and magnetars originate from a single continuous initial magnetic field distribution and in which magnetic inclination alignment plays a crucial role in shaping the observed population. Although additional effects, such as the recurrence time of bursts, are still required to explain the relative number of magnetars in samples extending to $\sim 10^5\,\mathrm{yr}$, magnetic inclination alignment is essential for linking pulsars and magnetars in the magnetic field distribution.

In the coming years, the full operation of next-generation radio facilities such as SKAO \citep{Grainge2017(SKA),Levin2025(SKA),Joshi2025(SKA)} and DSA-2000 \citep{Hallinan2019(DSA),Sherman2024(DSA)} is expected to increase the number of detected pulsars by one to two orders of magnitude, dramatically enlarging the population of neutron stars in the gap between pulsars and magnetars. This will make it possible to statistically investigate the inclination angle distribution and will provide a decisive observational basis for testing the evolutionary continuity between the populations.

\section*{Acknowledgment}

We thank Yong Gao, Anatoly Spitkovsky,  Gilad Sadeh, and Masaru Shibata for
helpful discussions. T.S. was supported by the JST SPRING, Grant Number JPMJSP2180. K.H.  was supported by the JST FOREST Program (JPMJFR2136) and the JSPS Grant-in-Aid for Scientific Research (23H01169, 23H04900). P.B. work was funded by a grant (no. 2024788) from the United States-Israel Binational Science Foundation (BSF), Jerusalem, Israel and by a grant (no. 1649/23) from the Israel Science Foundation.

\section*{DATA AVAILABILITY}
The data underlying this article will be shared on reasonable request
to the corresponding authors.

\bibliographystyle{mnras}
\bibliography{example}

@ARTICLE{Philippov2014,
       author = {{Philippov}, Alexander and {Tchekhovskoy}, Alexander and {Li}, Jason G.},
        title = "{Time evolution of pulsar obliquity angle from 3D simulations of magnetospheres}",
      journal = {\mnras},
         year = 2014,
        month = jul,
       volume = {441},
       number = {3},
        pages = {1879-1887},
          doi = {10.1093/mnras/stu591},
archivePrefix = {arXiv},
       eprint = {1311.1513},
 primaryClass = {astro-ph.HE},
       adsurl = {https://ui.adsabs.harvard.edu/abs/2014MNRAS.441.1879P}
}

@ARTICLE{Nakar2024Natur,
       author = {{Rodr{\'\i}guez}, {\'O}smar and {Nakar}, Ehud and {Maoz}, Dan},
        title = "{Stripped-envelope supernova light curves argue for central engine activity}",
      journal = {\nat},
         year = 2024,
        month = apr,
       volume = {628},
       number = {8009},
        pages = {733-735},
          doi = {10.1038/s41586-024-07262-x},
archivePrefix = {arXiv},
       eprint = {2404.10846},
 primaryClass = {astro-ph.HE},
       adsurl = {https://ui.adsabs.harvard.edu/abs/2024Natur.628..733R}
}

@ARTICLE{Tauris1998,
       author = {{Tauris}, T.~M. and {Manchester}, R.~N.},
        title = "{On the Evolution of Pulsar Beams}",
      journal = {\mnras},
         year = 1998,
        month = aug,
       volume = {298},
       number = {3},
        pages = {625-636},
          doi = {10.1046/j.1365-8711.1998.01369.x},
       adsurl = {https://ui.adsabs.harvard.edu/abs/1998MNRAS.298..625T}
}

@ARTICLE{Rankin1993,
       author = {{Rankin}, Joanna M.},
        title = "{Toward an Empirical Theory of Pulsar Emission. VI. The Geometry of the Conal Emission Region}",
      journal = {\apj},
         year = 1993,
        month = mar,
       volume = {405},
        pages = {285},
          doi = {10.1086/172361},
       adsurl = {https://ui.adsabs.harvard.edu/abs/1993ApJ...405..285R}
}

@ARTICLE{Gould1998,
       author = {{Gould}, D.~M. and {Lyne}, A.~G.},
        title = "{Multifrequency polarimetry of 300 radio pulsars}",
      journal = {\mnras},
         year = 1998,
        month = nov,
       volume = {301},
       number = {1},
        pages = {235-260},
          doi = {10.1046/j.1365-8711.1998.02018.x},
       adsurl = {https://ui.adsabs.harvard.edu/abs/1998MNRAS.301..235G}
}

@ARTICLE{Spitkovsky2006,
       author = {{Spitkovsky}, Anatoly},
        title = "{Time-dependent Force-free Pulsar Magnetospheres: Axisymmetric and Oblique Rotators}",
      journal = {\apjl},
         year = 2006,
        month = sep,
       volume = {648},
       number = {1},
        pages = {L51-L54},
          doi = {10.1086/507518},
archivePrefix = {arXiv},
       eprint = {astro-ph/0603147},
 primaryClass = {astro-ph},
       adsurl = {https://ui.adsabs.harvard.edu/abs/2006ApJ...648L..51S}
}

@ARTICLE{Goldreich1992,
       author = {{Goldreich}, Peter and {Reisenegger}, Andreas},
        title = "{Magnetic Field Decay in Isolated Neutron Stars}",
      journal = {\apj},
         year = 1992,
        month = aug,
       volume = {395},
        pages = {250},
          doi = {10.1086/171646},
       adsurl = {https://ui.adsabs.harvard.edu/abs/1992ApJ...395..250G}
}

@ARTICLE{Pardo-Araujo2026,
       author = {{Pardo-Araujo}, Celsa and {Rea}, Nanda and {Ronchi}, Michele and {Graber}, Vanessa},
        title = "{Magnetar fraction in Core-Collapse Supernovae}",
      journal = {arXiv e-prints},
         year = 2026,
        month = jan,
          eid = {arXiv:2601.16159},
        pages = {arXiv:2601.16159},
          doi = {10.48550/arXiv.2601.16159},
archivePrefix = {arXiv},
       eprint = {2601.16159},
 primaryClass = {astro-ph.HE},
       adsurl = {https://ui.adsabs.harvard.edu/abs/2026arXiv260116159P}
}

@ARTICLE{Kasen2010,
       author = {{Kasen}, Daniel and {Bildsten}, Lars},
        title = "{Supernova Light Curves Powered by Young Magnetars}",
      journal = {\apj},
         year = 2010,
        month = jul,
       volume = {717},
       number = {1},
        pages = {245-249},
          doi = {10.1088/0004-637X/717/1/245},
archivePrefix = {arXiv},
       eprint = {0911.0680},
 primaryClass = {astro-ph.HE},
       adsurl = {https://ui.adsabs.harvard.edu/abs/2010ApJ...717..245K}
}

@ARTICLE{Beniamini2023,
       author = {{Beniamini}, P. and {Wadiasingh}, Z. and {Hare}, J. and {Rajwade}, K.~M. and {Younes}, G. and {van der Horst}, A.~J.},
        title = "{Evidence for an abundant old population of Galactic ultra-long period magnetars and implications for fast radio bursts}",
      journal = {\mnras},
         year = 2023,
        month = apr,
       volume = {520},
       number = {2},
        pages = {1872-1894},
          doi = {10.1093/mnras/stad208},
archivePrefix = {arXiv},
       eprint = {2210.09323},
 primaryClass = {astro-ph.HE},
       adsurl = {https://ui.adsabs.harvard.edu/abs/2023MNRAS.520.1872B}
}

@ARTICLE{Men2025,
       author = {{Men}, Yunpeng and {McSweeney}, Sam and {Hurley-Walker}, Natasha and {Barr}, Ewan and {Stappers}, Ben},
        title = "{A highly magnetized long-period radio transient exhibiting unusual emission features}",
      journal = {Science Advances},
         year = 2025,
        month = jan,
       volume = {11},
       number = {3},
          eid = {eadp6351},
        pages = {eadp6351},
          doi = {10.1126/sciadv.adp6351},
archivePrefix = {arXiv},
       eprint = {2501.10528},
 primaryClass = {astro-ph.HE},
       adsurl = {https://ui.adsabs.harvard.edu/abs/2025SciA...11P6351M}
}

@ARTICLE{Cooper2024,
       author = {{Cooper}, A.~J. and {Wadiasingh}, Z.},
        title = "{Beyond the Rotational Deathline: Radio Emission from Ultra-long Period Magnetars}",
      journal = {\mnras},
         year = 2024,
        month = sep,
       volume = {533},
       number = {2},
        pages = {2133-2155},
          doi = {10.1093/mnras/stae1813},
archivePrefix = {arXiv},
       eprint = {2406.04135},
 primaryClass = {astro-ph.HE},
       adsurl = {https://ui.adsabs.harvard.edu/abs/2024MNRAS.533.2133C}
}

@ARTICLE{Usov1992,
       author = {{Usov}, V.~V.},
        title = "{Millisecond pulsars with extremely strong magnetic fields as a cosmological source of {\ensuremath{\gamma}}-ray bursts}",
      journal = {\nat},
         year = 1992,
        month = jun,
       volume = {357},
       number = {6378},
        pages = {472-474},
          doi = {10.1038/357472a0},
       adsurl = {https://ui.adsabs.harvard.edu/abs/1992Natur.357..472U}
}

@ARTICLE{Thompson2004,
       author = {{Thompson}, Todd A. and {Chang}, Philip and {Quataert}, Eliot},
        title = "{Magnetar Spin-Down, Hyperenergetic Supernovae, and Gamma-Ray Bursts}",
      journal = {\apj},
         year = 2004,
        month = aug,
       volume = {611},
       number = {1},
        pages = {380-393},
          doi = {10.1086/421969},
archivePrefix = {arXiv},
       eprint = {astro-ph/0401555},
 primaryClass = {astro-ph},
       adsurl = {https://ui.adsabs.harvard.edu/abs/2004ApJ...611..380T}
}

@ARTICLE{Bochenek2020,
       author = {{Bochenek}, C.~D. and {Ravi}, V. and {Belov}, K.~V. and {Hallinan}, G. and {Kocz}, J. and {Kulkarni}, S.~R. and {McKenna}, D.~L.},
        title = "{A fast radio burst associated with a Galactic magnetar}",
      journal = {\nat},
         year = 2020,
        month = nov,
       volume = {587},
       number = {7832},
        pages = {59-62},
          doi = {10.1038/s41586-020-2872-x},
archivePrefix = {arXiv},
       eprint = {2005.10828},
 primaryClass = {astro-ph.HE},
       adsurl = {https://ui.adsabs.harvard.edu/abs/2020Natur.587...59B}
}

@ARTICLE{Petroff2022,
       author = {{Petroff}, E. and {Hessels}, J.~W.~T. and {Lorimer}, D.~R.},
        title = "{Fast radio bursts at the dawn of the 2020s}",
      journal = {\aapr},
         year = 2022,
        month = dec,
       volume = {30},
       number = {1},
          eid = {2},
        pages = {2},
          doi = {10.1007/s00159-022-00139-w},
archivePrefix = {arXiv},
       eprint = {2107.10113},
 primaryClass = {astro-ph.HE},
       adsurl = {https://ui.adsabs.harvard.edu/abs/2022A&ARv..30....2P}
}

@ARTICLE{Weltevrede2008,
       author = {{Weltevrede}, Patrick and {Johnston}, Simon},
        title = "{The population of pulsars with interpulses and the implications for beam evolution}",
      journal = {\mnras},
         year = 2008,
        month = jul,
       volume = {387},
       number = {4},
        pages = {1755-1760},
          doi = {10.1111/j.1365-2966.2008.13382.x},
archivePrefix = {arXiv},
       eprint = {0804.4318},
 primaryClass = {astro-ph},
       adsurl = {https://ui.adsabs.harvard.edu/abs/2008MNRAS.387.1755W}
}

@ARTICLE{Beniamini2019,
       author = {{Beniamini}, Paz and {Hotokezaka}, Kenta and {van der Horst}, Alexander and {Kouveliotou}, Chryssa},
        title = "{Formation rates and evolution histories of magnetars}",
      journal = {\mnras},
         year = 2019,
        month = jul,
       volume = {487},
       number = {1},
        pages = {1426-1438},
          doi = {10.1093/mnras/stz1391},
archivePrefix = {arXiv},
       eprint = {1903.06718},
 primaryClass = {astro-ph.HE},
       adsurl = {https://ui.adsabs.harvard.edu/abs/2019MNRAS.487.1426B}
}

@ARTICLE{Shi2024,
       author = {{Shi}, Zhihong and {Ng}, C.-Y.},
        title = "{Pulsar Population Synthesis with Magnetorotational Evolution: Constraining the Decay of the Magnetic Field}",
      journal = {\apj},
         year = 2024,
        month = sep,
       volume = {972},
       number = {1},
          eid = {78},
        pages = {78},
          doi = {10.3847/1538-4357/ad5af8},
archivePrefix = {arXiv},
       eprint = {2408.14786},
 primaryClass = {astro-ph.HE},
       adsurl = {https://ui.adsabs.harvard.edu/abs/2024ApJ...972...78S}
}

@ARTICLE{Sautron2025,
       author = {{Sautron}, M. and {McEwen}, A.~E. and {Younes}, G. and {P{\'e}tri}, J. and {Beniamini}, P. and {Huppenkothen}, D.},
        title = "{The Galactic Population of Magnetars: A Simulation-based Inference Study}",
      journal = {\apj},
         year = 2025,
        month = jun,
       volume = {986},
       number = {1},
          eid = {88},
        pages = {88},
          doi = {10.3847/1538-4357/add0aa},
archivePrefix = {arXiv},
       eprint = {2503.11875},
 primaryClass = {astro-ph.HE},
       adsurl = {https://ui.adsabs.harvard.edu/abs/2025ApJ...986...88S}
}

@ARTICLE{Arzamasskiy2017,
       author = {{Arzamasskiy}, L.~I. and {Beskin}, V.~S. and {Pirov}, K.~K.},
        title = "{Statistics of interpulse radio pulsars: the key to solving the alignment/counter-alignment problem}",
      journal = {\mnras},
         year = 2017,
        month = apr,
       volume = {466},
       number = {2},
        pages = {2325-2336},
          doi = {10.1093/mnras/stw3139},
archivePrefix = {arXiv},
       eprint = {1612.04820},
 primaryClass = {astro-ph.HE},
       adsurl = {https://ui.adsabs.harvard.edu/abs/2017MNRAS.466.2325A}
}

@ARTICLE{Beskin1984,
       author = {{Beskin}, V.~S. and {Gurevich}, A.~V. and {Istomin}, Ia. N.},
        title = "{Spin-Down of Pulsars by the Current - Comparison of Theory with Observations}",
      journal = {\apss},
         year = 1984,
        month = jul,
       volume = {102},
       number = {2},
        pages = {301-326},
          doi = {10.1007/BF00650179},
       adsurl = {https://ui.adsabs.harvard.edu/abs/1984Ap&SS.102..301B}
}

@ARTICLE{Kniazev2024,
       author = {{Kniazev}, F.~A. and {Istomin}, A. Yu. and {Beskin}, V.~S.},
        title = "{Data From FAST and MeerKAT Surveys As a Test of Radio Pulsar Physics}",
      journal = {Astronomy Letters},
         year = 2024,
        month = dec,
       volume = {50},
       number = {12},
        pages = {821-829},
          doi = {10.1134/S1063773725700148},
archivePrefix = {arXiv},
       eprint = {2506.12423},
 primaryClass = {astro-ph.HE},
       adsurl = {https://ui.adsabs.harvard.edu/abs/2024AstL...50..821K}
}

@BOOK{Beskin1993,
       author = {{Beskin}, V.~S. and {Gurevich}, Aleksandr V. and {Istomin}, Ya. N.},
        title = "{Physics of the pulsar magnetosphere}",
         year = 1993,
       adsurl = {https://ui.adsabs.harvard.edu/abs/1993ppm..book.....B}
}

@ARTICLE{Rookyard2015a,
       author = {{Rookyard}, S.~C. and {Weltevrede}, P. and {Johnston}, S.},
        title = "{Investigating the magnetic inclination angle distribution of {\ensuremath{\gamma}}-ray-loud radio pulsars}",
      journal = {\mnras},
         year = 2015,
        month = feb,
       volume = {446},
       number = {4},
        pages = {3356-3366},
          doi = {10.1093/mnras/stu2083},
archivePrefix = {arXiv},
       eprint = {1410.3310},
 primaryClass = {astro-ph.HE},
       adsurl = {https://ui.adsabs.harvard.edu/abs/2015MNRAS.446.3356R}
}

@ARTICLE{Rookyard2015b,
       author = {{Rookyard}, S.~C. and {Weltevrede}, P. and {Johnston}, S.},
        title = "{Constraints on viewing geometries from radio observations of {\ensuremath{\gamma}}-ray-loud pulsars using a novel method}",
      journal = {\mnras},
         year = 2015,
        month = feb,
       volume = {446},
       number = {4},
        pages = {3367-3388},
          doi = {10.1093/mnras/stu2236},
archivePrefix = {arXiv},
       eprint = {1410.3294},
 primaryClass = {astro-ph.HE},
       adsurl = {https://ui.adsabs.harvard.edu/abs/2015MNRAS.446.3367R}
}

@ARTICLE{Lyne2015,
       author = {{Lyne}, A.~G. and {Jordan}, C.~A. and {Graham-Smith}, F. and {Espinoza}, C.~M. and {Stappers}, B.~W. and {Weltevrede}, P.},
        title = "{45 years of rotation of the Crab pulsar}",
      journal = {\mnras},
         year = 2015,
        month = jan,
       volume = {446},
       number = {1},
        pages = {857-864},
          doi = {10.1093/mnras/stu2118},
archivePrefix = {arXiv},
       eprint = {1410.0886},
 primaryClass = {astro-ph.HE},
       adsurl = {https://ui.adsabs.harvard.edu/abs/2015MNRAS.446..857L}
}

@ARTICLE{Zanazzi2015,
       author = {{Zanazzi}, J.~J. and {Lai}, Dong},
        title = "{Electromagnetic torques, precession and evolution of magnetic inclination of pulsars}",
      journal = {\mnras},
         year = 2015,
        month = jul,
       volume = {451},
       number = {1},
        pages = {695-704},
          doi = {10.1093/mnras/stv955},
archivePrefix = {arXiv},
       eprint = {1503.01486},
 primaryClass = {astro-ph.SR},
       adsurl = {https://ui.adsabs.harvard.edu/abs/2015MNRAS.451..695Z}
}

@ARTICLE{Igoshev2020,
       author = {{Igoshev}, Andrei P. and {Popov}, Sergei B.},
        title = "{Braking indices of young radio pulsars: theoretical perspective}",
      journal = {\mnras},
         year = 2020,
        month = dec,
       volume = {499},
       number = {2},
        pages = {2826-2835},
          doi = {10.1093/mnras/staa3070},
archivePrefix = {arXiv},
       eprint = {2008.11737},
 primaryClass = {astro-ph.HE},
       adsurl = {https://ui.adsabs.harvard.edu/abs/2020MNRAS.499.2826I}
}

@ARTICLE{Desvignes2024,
       author = {{Desvignes}, Gregory and {Weltevrede}, Patrick and {Gao}, Yong and {Jones}, David Ian and {Kramer}, Michael and {Caleb}, Manisha and {Karuppusamy}, Ramesh and {Levin}, Lina and {Liu}, Kuo and {Lyne}, Andrew G. and {Shao}, Lijing and {Stappers}, Ben and {P{\'e}tri}, J{\'e}r{\^o}me},
        title = "{A freely precessing magnetar following an X-ray outburst}",
      journal = {Nature Astronomy},
         year = 2024,
        month = may,
       volume = {8},
        pages = {617-627},
          doi = {10.1038/s41550-024-02226-7},
       adsurl = {https://ui.adsabs.harvard.edu/abs/2024NatAs...8..617D}
}

@ARTICLE{Camilo2006,
       author = {{Camilo}, Fernando and {Ransom}, Scott M. and {Halpern}, Jules P. and {Reynolds}, John and {Helfand}, David J. and {Zimmerman}, Neil and {Sarkissian}, John},
        title = "{Transient pulsed radio emission from a magnetar}",
      journal = {\nat},
         year = 2006,
        month = aug,
       volume = {442},
       number = {7105},
        pages = {892-895},
          doi = {10.1038/nature04986},
archivePrefix = {arXiv},
       eprint = {astro-ph/0605429},
 primaryClass = {astro-ph},
       adsurl = {https://ui.adsabs.harvard.edu/abs/2006Natur.442..892C}
}

@ARTICLE{Kramer2007,
       author = {{Kramer}, M. and {Stappers}, B.~W. and {Jessner}, A. and {Lyne}, A.~G. and {Jordan}, C.~A.},
        title = "{Polarized radio emission from a magnetar}",
      journal = {\mnras},
         year = 2007,
        month = may,
       volume = {377},
       number = {1},
        pages = {107-119},
          doi = {10.1111/j.1365-2966.2007.11622.x},
archivePrefix = {arXiv},
       eprint = {astro-ph/0702365},
 primaryClass = {astro-ph},
       adsurl = {https://ui.adsabs.harvard.edu/abs/2007MNRAS.377..107K}
}

@ARTICLE{Camilo2008,
       author = {{Camilo}, F. and {Reynolds}, J. and {Johnston}, S. and {Halpern}, J.~P. and {Ransom}, S.~M.},
        title = "{The Magnetar 1E 1547.0-5408: Radio Spectrum, Polarimetry, and Timing}",
      journal = {\apj},
         year = 2008,
        month = may,
       volume = {679},
       number = {1},
        pages = {681-686},
          doi = {10.1086/587054},
archivePrefix = {arXiv},
       eprint = {0802.0494},
 primaryClass = {astro-ph},
       adsurl = {https://ui.adsabs.harvard.edu/abs/2008ApJ...679..681C}
}

@ARTICLE{Camilo2007,
       author = {{Camilo}, F. and {Ransom}, S.~M. and {Halpern}, J.~P. and {Reynolds}, J.},
        title = "{1E 1547.0-5408: A Radio-emitting Magnetar with a Rotation Period of 2 Seconds}",
      journal = {\apjl},
         year = 2007,
        month = sep,
       volume = {666},
       number = {2},
        pages = {L93-L96},
          doi = {10.1086/521826},
archivePrefix = {arXiv},
       eprint = {0708.0002},
 primaryClass = {astro-ph},
       adsurl = {https://ui.adsabs.harvard.edu/abs/2007ApJ...666L..93C}
}

@ARTICLE{Levin2012,
       author = {{Levin}, L. and {Bailes}, M. and {Bates}, S.~D. and {Bhat}, N.~D.~R. and {Burgay}, M. and {Burke-Spolaor}, S. and {D'Amico}, N. and {Johnston}, S. and {Keith}, M.~J. and {Kramer}, M. and {Milia}, S. and {Possenti}, A. and {Stappers}, B. and {van Straten}, W.},
        title = "{Radio emission evolution, polarimetry and multifrequency single pulse analysis of the radio magnetar PSR J1622-4950}",
      journal = {\mnras},
         year = 2012,
        month = may,
       volume = {422},
       number = {3},
        pages = {2489-2500},
          doi = {10.1111/j.1365-2966.2012.20807.x},
archivePrefix = {arXiv},
       eprint = {1204.2045},
 primaryClass = {astro-ph.HE},
       adsurl = {https://ui.adsabs.harvard.edu/abs/2012MNRAS.422.2489L}
}

@ARTICLE{Lower2020,
       author = {{Lower}, Marcus E. and {Shannon}, Ryan M. and {Johnston}, Simon and {Bailes}, Matthew},
        title = "{Spectropolarimetric Properties of Swift J1818.0-1607: A 1.4 s Radio Magnetar}",
      journal = {\apjl},
         year = 2020,
        month = jun,
       volume = {896},
       number = {2},
          eid = {L37},
        pages = {L37},
          doi = {10.3847/2041-8213/ab9898},
archivePrefix = {arXiv},
       eprint = {2004.11522},
 primaryClass = {astro-ph.HE},
       adsurl = {https://ui.adsabs.harvard.edu/abs/2020ApJ...896L..37L}
}

@ARTICLE{Eksi2016,
       author = {{Ek{\c{s}}i}, K.~Y. and {Anda{\c{c}}}, I.~C. and {{\c{C}}{\i}k{\i}nto{\u{g}}lu}, S. and {G{\"u}gercino{\u{g}}lu}, E. and {Vahdat Motlagh}, A. and {K{\i}z{\i}ltan}, B.},
        title = "{The Inclination Angle and Evolution of the Braking Index of Pulsars with Plasma-filled Magnetosphere: Application to the High Braking Index of PSR J1640-4631}",
      journal = {\apj},
         year = 2016,
        month = may,
       volume = {823},
       number = {1},
          eid = {34},
        pages = {34},
          doi = {10.3847/0004-637X/823/1/34},
archivePrefix = {arXiv},
       eprint = {1603.01487},
 primaryClass = {astro-ph.HE},
       adsurl = {https://ui.adsabs.harvard.edu/abs/2016ApJ...823...34E}
}

@ARTICLE{Zeng2026,
       author = {{Zeng}, Shuzhe and {Philippov}, Alexander and {Juno}, James and {Beloborodov}, Andrei M. and {Popova}, Elena},
        title = "{Origin of Pulsed Radio Emission from Magnetars}",
      journal = {\apjl},
         year = 2026,
        month = jan,
       volume = {996},
       number = {2},
          eid = {L20},
        pages = {L20},
          doi = {10.3847/2041-8213/ae2ade},
archivePrefix = {arXiv},
       eprint = {2509.13419},
 primaryClass = {astro-ph.HE},
       adsurl = {https://ui.adsabs.harvard.edu/abs/2026ApJ...996L..20Z}
}

@ARTICLE{Radhakrishnan1969,
       author = {{Radhakrishnan}, V. and {Cooke}, D.~J.},
        title = "{Magnetic Poles and the Polarization Structure of Pulsar Radiation}",
      journal = {\aplett},
         year = 1969,
        month = jan,
       volume = {3},
        pages = {225},
       adsurl = {https://ui.adsabs.harvard.edu/abs/1969ApL.....3..225R}
}

@ARTICLE{Li2026,
       author = {{Li}, B.~P. and {Gao}, Z.~F. and {Ma}, W.~Q. and {Zhang}, W.~F. and {Cheng}, Q.},
        title = "{Studies on the Spin and Magnetic Inclination Evolution of Magnetars Swift J1834.9─0846 under Wind Braking}",
      journal = {\apj},
         year = 2026,
        month = mar,
       volume = {999},
       number = {2},
          eid = {262},
        pages = {262},
          doi = {10.3847/1538-4357/ae3bcf},
archivePrefix = {arXiv},
       eprint = {2602.06615},
 primaryClass = {astro-ph.HE},
       adsurl = {https://ui.adsabs.harvard.edu/abs/2026ApJ...999..262L}
}

@ARTICLE{Camilo2018,
       author = {{Camilo}, F. and {Scholz}, P. and {Serylak}, M. and {Buchner}, S. and {Merryfield}, M. and {Kaspi}, V.~M. and {Archibald}, R.~F. and {Bailes}, M. and {Jameson}, A. and {van Straten}, W. and {Sarkissian}, J. and {Reynolds}, J.~E. and {Johnston}, S. and {Hobbs}, G. and {Abbott}, T.~D. and {Adam}, R.~M. and {Adams}, G.~B. and {Alberts}, T. and {Andreas}, R. and {Asad}, K.~M.~B. and {Baker}, D.~E. and {Baloyi}, T. and {Bauermeister}, E.~F. and {Baxana}, T. and {Bennett}, T.~G.~H. and {Bernardi}, G. and {Booisen}, D. and {Booth}, R.~S. and {Botha}, D.~H. and {Boyana}, L. and {Brederode}, L.~R.~S. and {Burger}, J.~P. and {Cheetham}, T. and {Conradie}, J. and {Conradie}, J.~P. and {Davidson}, D.~B. and {De Bruin}, G. and {de Swardt}, B. and {de Villiers}, C. and {de Villiers}, D.~I.~L. and {de Villiers}, M.~S. and {de Villiers}, W. and {De Waal}, C. and {Dikgale}, M.~A. and {du Toit}, G. and {du Toit}, L.~J. and {Esterhuyse}, S.~W.~P. and {Fanaroff}, B. and {Fataar}, S. and {Foley}, A.~R. and {Foster}, G. and {Fourie}, D. and {Gamatham}, R. and {Gatsi}, T. and {Geschke}, R. and {Goedhart}, S. and {Grobler}, T.~L. and {Gumede}, S.~C. and {Hlakola}, M.~J. and {Hokwana}, A. and {Hoorn}, D.~H. and {Horn}, D. and {Horrell}, J. and {Hugo}, B. and {Isaacson}, A. and {Jacobs}, O. and {Jansen van Rensburg}, J.~P. and {Jonas}, J.~L. and {Jordaan}, B. and {Joubert}, A. and {Joubert}, F. and {J{\'o}zsa}, G.~I.~G. and {Julie}, R. and {Julius}, C.~C. and {Kapp}, F. and {Karastergiou}, A. and {Karels}, F. and {Kariseb}, M. and {Karuppusamy}, R. and {Kasper}, V. and {Knox-Davies}, E.~C. and {Koch}, D. and {Kotz{\'e}}, P.~P.~A. and {Krebs}, A. and {Kriek}, N. and {Kriel}, H. and {Kusel}, T. and {Lamoor}, S. and {Lehmensiek}, R. and {Liebenberg}, D. and {Liebenberg}, I. and {Lord}, R.~T. and {Lunsky}, B. and {Mabombo}, N. and {Macdonald}, T. and {Macfarlane}, P. and {Madisa}, K. and {Mafhungo}, L. and {Magnus}, L.~G. and {Magozore}, C. and {Mahgoub}, O. and {Main}, J.~P.~L. and {Makhathini}, S. and {Malan}, J.~A. and {Malgas}, P. and {Manley}, J.~R. and {Manzini}, M. and {Marais}, L. and {Marais}, N. and {Marais}, S.~J. and {Maree}, M. and {Martens}, A. and {Matshawule}, S.~D. and {Matthysen}, N. and {Mauch}, T. and {McNally}, L.~D. and {Merry}, B. and {Millenaar}, R.~P. and {Mjikelo}, C. and {Mkhabela}, N. and {Mnyandu}, N. and {Moeng}, I.~T. and {Mokone}, O.~J. and {Monama}, T.~E. and {Montshiwa}, K. and {Moss}, V. and {Mphego}, M. and {New}, W. and {Ngcebetsha}, B. and {Ngoasheng}, K. and {Niehaus}, H. and {Ntuli}, P. and {Nzama}, A. and {Obies}, F. and {Obrocka}, M. and {Ockards}, M.~T. and {Olyn}, C. and {Oozeer}, N. and {Otto}, A.~J. and {Padayachee}, Y. and {Passmoor}, S. and {Patel}, A.~A. and {Paula}, S. and {Peens-Hough}, A. and {Pholoholo}, B. and {Prozesky}, P. and {Rakoma}, S. and {Ramaila}, A.~J.~T. and {Rammala}, I. and {Ramudzuli}, Z.~R. and {Rasivhaga}, M. and {Ratcliffe}, S. and {Reader}, H.~C. and {Renil}, R. and {Richter}, L. and {Robyntjies}, A. and {Rosekrans}, D. and {Rust}, A. and {Salie}, S. and {Sambu}, N. and {Schollar}, C.~T.~G. and {Schwardt}, L. and {Seranyane}, S. and {Sethosa}, G. and {Sharpe}, C. and {Siebrits}, R. and {Sirothia}, S.~K. and {Slabber}, M.~J. and {Smirnov}, O. and {Smith}, S. and {Sofeya}, L. and {Songqumase}, N. and {Spann}, R. and {Stappers}, B. and {Steyn}, D. and {Steyn}, T.~J. and {Strong}, R. and {Struthers}, A. and {Stuart}, C. and {Sunnylall}, P. and {Swart}, P.~S. and {Taljaard}, B. and {Tasse}, C. and {Taylor}, G. and {Theron}, I.~P. and {Thondikulam}, V. and {Thorat}, K. and {Tiplady}, A. and {Toruvanda}, O. and {van Aardt}, J. and {van Balla}, T. and {van den Heever}, L. and {van der Byl}, A. and {van der Merwe}, C. and {van der Merwe}, P. and {van Niekerk}, P.~C. and {van Rooyen}, R. and {van Staden}, J.~P. and {van Tonder}, V. and {van Wyk}, R.},
        title = "{Revival of the Magnetar PSR J1622-4950: Observations with MeerKAT, Parkes, XMM-Newton, Swift, Chandra, and NuSTAR}",
      journal = {\apj},
         year = 2018,
        month = apr,
       volume = {856},
       number = {2},
          eid = {180},
        pages = {180},
          doi = {10.3847/1538-4357/aab35a},
archivePrefix = {arXiv},
       eprint = {1804.01933},
 primaryClass = {astro-ph.HE},
       adsurl = {https://ui.adsabs.harvard.edu/abs/2018ApJ...856..180C}
}

@ARTICLE{Camilo2007XTE,
       author = {{Camilo}, F. and {Cognard}, I. and {Ransom}, S.~M. and {Halpern}, J.~P. and {Reynolds}, J. and {Zimmerman}, N. and {Gotthelf}, E.~V. and {Helfand}, D.~J. and {Demorest}, P. and {Theureau}, G. and {Backer}, D.~C.},
        title = "{The Magnetar XTE J1810-197: Variations in Torque, Radio Flux Density, and Pulse Profile Morphology}",
      journal = {\apj},
         year = 2007,
        month = jul,
       volume = {663},
       number = {1},
        pages = {497-504},
          doi = {10.1086/518226},
archivePrefix = {arXiv},
       eprint = {astro-ph/0610685},
 primaryClass = {astro-ph},
       adsurl = {https://ui.adsabs.harvard.edu/abs/2007ApJ...663..497C}
}

@ARTICLE{Lower2021,
       author = {{Lower}, M.~E. and {Johnston}, S. and {Shannon}, R.~M. and {Bailes}, M. and {Camilo}, F.},
        title = "{The dynamic magnetosphere of Swift J1818.0-1607}",
      journal = {\mnras},
         year = 2021,
        month = mar,
       volume = {502},
       number = {1},
        pages = {127-139},
          doi = {10.1093/mnras/staa3789},
archivePrefix = {arXiv},
       eprint = {2011.12463},
 primaryClass = {astro-ph.HE},
       adsurl = {https://ui.adsabs.harvard.edu/abs/2021MNRAS.502..127L}
}

@ARTICLE{delima2020,
       author = {{de Lima}, Rafael C.~R. and {Coelho}, Jaziel G. and {Pereira}, Jonas P. and {Rodrigues}, Claudia V. and {Rueda}, Jorge A.},
        title = "{Evidence for a Multipolar Magnetic Field in SGR J1745-2900 from X-Ray Light-curve Analysis}",
      journal = {\apj},
         year = 2020,
        month = feb,
       volume = {889},
       number = {2},
          eid = {165},
        pages = {165},
          doi = {10.3847/1538-4357/ab65f4},
archivePrefix = {arXiv},
       eprint = {1912.12336},
 primaryClass = {astro-ph.SR},
       adsurl = {https://ui.adsabs.harvard.edu/abs/2020ApJ...889..165D}
}

@ARTICLE{Zane2023,
       author = {{Zane}, Silvia and {Taverna}, Roberto and {Gonz{\'a}lez-Caniulef}, Denis and {Muleri}, Fabio and {Turolla}, Roberto and {Heyl}, Jeremy and {Uchiyama}, Keisuke and {Ng}, Mason and {Tamagawa}, Toru and {Caiazzo}, Ilaria and {Di Lalla}, Niccol{\`o} and {Marshall}, Herman L. and {Bachetti}, Matteo and {La Monaca}, Fabio and {Gau}, Ephraim and {Di Marco}, Alessandro and {Baldini}, Luca and {Negro}, Michela and {Omodei}, Nicola and {Rankin}, John and {Matt}, Giorgio and {Pavlov}, George G. and {Kitaguchi}, Takao and {Krawczynski}, Henric and {Kislat}, Fabian and {Kelly}, Ruth and {Agudo}, Iv{\'a}n and {Antonelli}, Lucio A. and {Baumgartner}, Wayne H. and {Bellazzini}, Ronaldo and {Bianchi}, Stefano and {Bongiorno}, Stephen D. and {Bonino}, Raffaella and {Brez}, Alessandro and {Bucciantini}, Niccol{\`o} and {Capitanio}, Fiamma and {Castellano}, Simone and {Cavazzuti}, Elisabetta and {Chen}, Chieng-Ting and {Ciprini}, Stefano and {Costa}, Enrico and {De Rosa}, Alessandra and {Del Monte}, Ettore and {Di Gesu}, Laura and {Donnarumma}, Immacolata and {Doroshenko}, Victor and {Dov{\v{c}}iak}, Michal and {Ehlert}, Steven R. and {Enoto}, Teruaki and {Evangelista}, Yuri and {Fabiani}, Sergio and {Ferrazzoli}, Riccardo and {Garcia}, Javier A. and {Gunji}, Shuichi and {Hayashida}, Kiyoshi and {Iwakiri}, Wataru and {Jorstad}, Svetlana G. and {Kaaret}, Philip and {Karas}, Vladimir and {Kolodziejczak}, Jeffery J. and {Latronico}, Luca and {Liodakis}, Ioannis and {Maldera}, Simone and {Manfreda}, Alberto and {Marin}, Fr{\'e}d{\'e}ric and {Marinucci}, Andrea and {Marscher}, Alan P. and {Massaro}, Francesco and {Mitsuishi}, Ikuyuki and {Mizuno}, Tsunefumi and {Ng}, C.-Y. and {O'Dell}, Stephen L. and {Oppedisano}, Chiara and {Papitto}, Alessandro and {Peirson}, Abel L. and {Perri}, Matteo and {Pesce-Rollins}, Melissa and {Petrucci}, Pierre-Olivier and {Pilia}, Maura and {Possenti}, Andrea and {Poutanen}, Juri and {Puccetti}, Simonetta and {Ramsey}, Brian D. and {Ratheesh}, Ajay and {Roberts}, Oliver J. and {Romani}, Roger W. and {Sgr{\'o}}, Carmelo and {Slane}, Patrick and {Soffitta}, Paolo and {Spandre}, Gloria and {Swartz}, Douglas A. and {Tavecchio}, Fabrizio and {Tawara}, Yuzuru and {Tennant}, Allyn F. and {Thomas}, Nicholas E. and {Tombesi}, Francesco and {Trois}, Alessio and {Tsygankov}, Sergey S. and {Vink}, Jacco and {Weisskopf}, Martin C. and {Wu}, Kinwah and {Xie}, Fei},
        title = "{A Strong X-Ray Polarization Signal from the Magnetar 1RXS J170849.0-400910}",
      journal = {\apjl},
         year = 2023,
        month = feb,
       volume = {944},
       number = {2},
          eid = {L27},
        pages = {L27},
          doi = {10.3847/2041-8213/acb703},
archivePrefix = {arXiv},
       eprint = {2301.12919},
 primaryClass = {astro-ph.HE},
       adsurl = {https://ui.adsabs.harvard.edu/abs/2023ApJ...944L..27Z}
}

@ARTICLE{Heyl2024,
       author = {{Heyl}, Jeremy and {Taverna}, Roberto and {Turolla}, Roberto and {Israel}, Gian Luca and {Ng}, Mason and {K{\i}rm{\i}z{\i}bayrak}, Demet and {Gonz{\'a}lez-Caniulef}, Denis and {Caiazzo}, Ilaria and {Zane}, Silvia and {Ehlert}, Steven R. and {Negro}, Michela and {Agudo}, Iv{\'a}n and {Antonelli}, Lucio Angelo and {Bachetti}, Matteo and {Baldini}, Luca and {Baumgartner}, Wayne H. and {Bellazzini}, Ronaldo and {Bianchi}, Stefano and {Bongiorno}, Stephen D. and {Bonino}, Raffaella and {Brez}, Alessandro and {Bucciantini}, Niccol{\`o} and {Capitanio}, Fiamma and {Castellano}, Simone and {Cavazzuti}, Elisabetta and {Chen}, Chien-Ting and {Ciprini}, Stefano and {Costa}, Enrico and {De Rosa}, Alessandra and {Del Monte}, Ettore and {Di Gesu}, Laura and {Di Lalla}, Niccol{\`o} and {Di Marco}, Alessandro and {Donnarumma}, Immacolata and {Doroshenko}, Victor and {Dov{\v{c}}iak}, Michal and {Enoto}, Teruaki and {Evangelista}, Yuri and {Fabiani}, Sergio and {Ferrazzoli}, Riccardo and {Garcia}, Javier A. and {Gunji}, Shuichi and {Hayashida}, Kiyoshi and {Iwakiri}, Wataru and {Jorstad}, Svetlana G. and {Kaaret}, Philip and {Karas}, Vladimir and {Kislat}, Fabian and {Kitaguchi}, Takao and {Kolodziejczak}, Jeffery J. and {Krawczynski}, Henric and {La Monaca}, Fabio and {Latronico}, Luca and {Liodakis}, Ioannis and {Maldera}, Simone and {Manfreda}, Alberto and {Marin}, Fr{\'e}d{\'e}ric and {Marinucci}, Andrea and {Marscher}, Alan P. and {Marshall}, Herman L. and {Massaro}, Francesco and {Matt}, Giorgio and {Mitsuishi}, Ikuyuki and {Mizuno}, Tsunefumi and {Muleri}, Fabio and {Ng}, C.-Y. and {O'Dell}, Stephen L. and {Omodei}, Nicola and {Oppedisano}, Chiara and {Papitto}, Alessandro and {Pavlov}, George G. and {Peirson}, Abel Lawrence and {Perri}, Matteo and {Pesce-Rollins}, Melissa and {Petrucci}, Pierre-Olivier and {Pilia}, Maura and {Possenti}, Andrea and {Poutanen}, Juri and {Puccetti}, Simonetta and {Ramsey}, Brian D. and {Rankin}, John and {Ratheesh}, Ajay and {Roberts}, Oliver J. and {Romani}, Roger W. and {Sgr{\`o}}, Carmelo and {Slane}, Patrick and {Soffitta}, Paolo and {Spandre}, Gloria and {Swartz}, Douglas A. and {Tamagawa}, Toru and {Tavecchio}, Fabrizio and {Tawara}, Yuzuru and {Tennant}, Allyn F. and {Thomas}, Nicholas E. and {Tombesi}, Francesco and {Trois}, Alessio and {Tsygankov}, Sergey S. and {Vink}, Jacco and {Weisskopf}, Martin C. and {Wu}, Kinwah and {Xie}, Fei},
        title = "{The detection of polarized X-ray emission from the magnetar 1E 2259+586}",
      journal = {\mnras},
         year = 2024,
        month = feb,
       volume = {527},
       number = {4},
        pages = {12219-12231},
          doi = {10.1093/mnras/stad3680},
archivePrefix = {arXiv},
       eprint = {2311.03637},
 primaryClass = {astro-ph.HE},
       adsurl = {https://ui.adsabs.harvard.edu/abs/2024MNRAS.52712219H}
}

@ARTICLE{Zhu2023,
       author = {{Zhu}, Weiwei and {Xu}, Heng and {Zhou}, Dejiang and {Lin}, Lin and {Wang}, Bojun and {Wang}, Pei and {Zhang}, Chunfeng and {Niu}, Jiarui and {Chen}, Yutong and {Li}, Chengkui and {Meng}, Lingqi and {Lee}, Kejia and {Zhang}, Bing and {Feng}, Yi and {Ge}, Mingyu and {G{\"o}{\u{g}}{\"u}{\textcommabelow s}}, Ersin and {Guan}, Xing and {Han}, Jinlin and {Jiang}, Jinchen and {Jiang}, Peng and {Kouveliotou}, Chryssa and {Li}, Di and {Miao}, Chenchen and {Miao}, Xueli and {Men}, Yunpeng and {Niu}, Chenghui and {Wang}, Weiyang and {Wang}, Zhengli and {Xu}, Jiangwei and {Xu}, Renxin and {Xue}, Mengyao and {Yang}, Yuanpei and {Yu}, Wenfei and {Yuan}, Mao and {Yue}, Youling and {Zhang}, Shuangnan and {Zhang}, Yongkun},
        title = "{A radio pulsar phase from SGR J1935+2154 provides clues to the magnetar FRB mechanism}",
      journal = {Science Advances},
         year = 2023,
        month = jul,
       volume = {9},
       number = {30},
          eid = {eadf6198},
        pages = {eadf6198},
          doi = {10.1126/sciadv.adf6198},
archivePrefix = {arXiv},
       eprint = {2307.16124},
 primaryClass = {astro-ph.HE},
       adsurl = {https://ui.adsabs.harvard.edu/abs/2023SciA....9F6198Z}
}

@ARTICLE{Beniamini2025can,
       author = {{Beniamini}, Paz and {Kumar}, Pawan},
        title = "{Can Repeating and Nonrepeating Fast Radio Bursts Be Drawn from the Same Population?}",
      journal = {\apj},
         year = 2025,
        month = nov,
       volume = {993},
       number = {1},
          eid = {37},
        pages = {37},
          doi = {10.3847/1538-4357/ae0712},
archivePrefix = {arXiv},
       eprint = {2506.09138},
 primaryClass = {astro-ph.HE},
       adsurl = {https://ui.adsabs.harvard.edu/abs/2025ApJ...993...37B}
}

@ARTICLE{Beniamini2025role,
       author = {{Beniamini}, Paz and {Kumar}, Pawan},
        title = "{The Role of Magnetic and Rotation Axis Alignment in Driving Fast Radio Burst Phenomenology}",
      journal = {\apj},
         year = 2025,
        month = mar,
       volume = {982},
       number = {1},
          eid = {45},
        pages = {45},
          doi = {10.3847/1538-4357/adb8e6},
archivePrefix = {arXiv},
       eprint = {2410.19043},
 primaryClass = {astro-ph.HE},
       adsurl = {https://ui.adsabs.harvard.edu/abs/2025ApJ...982...45B}
}

@ARTICLE{Mereghetti2020,
       author = {{Mereghetti}, S. and {Savchenko}, V. and {Ferrigno}, C. and {G{\"o}tz}, D. and {Rigoselli}, M. and {Tiengo}, A. and {Bazzano}, A. and {Bozzo}, E. and {Coleiro}, A. and {Courvoisier}, T.~J.-L. and {Doyle}, M. and {Goldwurm}, A. and {Hanlon}, L. and {Jourdain}, E. and {von Kienlin}, A. and {Lutovinov}, A. and {Martin-Carrillo}, A. and {Molkov}, S. and {Natalucci}, L. and {Onori}, F. and {Panessa}, F. and {Rodi}, J. and {Rodriguez}, J. and {S{\'a}nchez-Fern{\'a}ndez}, C. and {Sunyaev}, R. and {Ubertini}, P.},
        title = "{INTEGRAL Discovery of a Burst with Associated Radio Emission from the Magnetar SGR 1935+2154}",
      journal = {\apjl},
         year = 2020,
        month = aug,
       volume = {898},
       number = {2},
          eid = {L29},
        pages = {L29},
          doi = {10.3847/2041-8213/aba2cf},
archivePrefix = {arXiv},
       eprint = {2005.06335},
 primaryClass = {astro-ph.HE},
       adsurl = {https://ui.adsabs.harvard.edu/abs/2020ApJ...898L..29M}
}

@ARTICLE{Gotthelf2014,
       author = {{Gotthelf}, E.~V. and {Tomsick}, J.~A. and {Halpern}, J.~P. and {Gelfand}, J.~D. and {Harrison}, F.~A. and {Boggs}, S.~E. and {Christensen}, F.~E. and {Craig}, W.~W. and {Hailey}, J.~C. and {Kaspi}, V.~M. and {Stern}, D.~K. and {Zhang}, W.~W.},
        title = "{NuSTAR Discovery of a Young, Energetic Pulsar Associated with the Luminous Gamma-Ray Source HESS J1640-465}",
      journal = {\apj},
         year = 2014,
        month = jun,
       volume = {788},
       number = {2},
          eid = {155},
        pages = {155},
          doi = {10.1088/0004-637X/788/2/155},
archivePrefix = {arXiv},
       eprint = {1405.0465},
 primaryClass = {astro-ph.HE},
       adsurl = {https://ui.adsabs.harvard.edu/abs/2014ApJ...788..155G}
}

@ARTICLE{Olausen2014,
       author = {{Olausen}, S.~A. and {Kaspi}, V.~M.},
        title = "{The McGill Magnetar Catalog}",
      journal = {\apjs},
         year = 2014,
        month = may,
       volume = {212},
       number = {1},
          eid = {6},
        pages = {6},
          doi = {10.1088/0067-0049/212/1/6},
archivePrefix = {arXiv},
       eprint = {1309.4167},
 primaryClass = {astro-ph.HE},
       adsurl = {https://ui.adsabs.harvard.edu/abs/2014ApJS..212....6O}
}

@ARTICLE{Beloborodov2013,
       author = {{Beloborodov}, Andrei M.},
        title = "{Electron-Positron Flows around Magnetars}",
      journal = {\apj},
         year = 2013,
        month = nov,
       volume = {777},
       number = {2},
          eid = {114},
        pages = {114},
          doi = {10.1088/0004-637X/777/2/114},
archivePrefix = {arXiv},
       eprint = {1209.4063},
 primaryClass = {astro-ph.HE},
       adsurl = {https://ui.adsabs.harvard.edu/abs/2013ApJ...777..114B}
}

@ARTICLE{Beniamini2025MGF,
       author = {{Beniamini}, Paz and {Wadiasingh}, Zorawar and {Trigg}, Aaron and {Chirenti}, Cecilia and {Burns}, Eric and {Younes}, George and {Negro}, Michela and {Granot}, Jonathan},
        title = "{Extragalactic Magnetar Giant Flares: Population Implications, Rates, and Prospects for Gamma-Rays, Gravitational Waves, and Neutrinos}",
      journal = {\apj},
         year = 2025,
        month = feb,
       volume = {980},
       number = {2},
          eid = {211},
        pages = {211},
          doi = {10.3847/1538-4357/ada947},
archivePrefix = {arXiv},
       eprint = {2411.16846},
 primaryClass = {astro-ph.HE},
       adsurl = {https://ui.adsabs.harvard.edu/abs/2025ApJ...980..211B}
}

@ARTICLE{Dai2019,
       author = {{Dai}, Shi and {Lower}, Marcus E. and {Bailes}, Matthew and {Camilo}, Fernando and {Halpern}, Jules P. and {Johnston}, Simon and {Kerr}, Matthew and {Reynolds}, John and {Sarkissian}, John and {Scholz}, Paul},
        title = "{Wideband Polarized Radio Emission from the Newly Revived Magnetar XTE J1810-197}",
      journal = {\apjl},
         year = 2019,
        month = apr,
       volume = {874},
       number = {2},
          eid = {L14},
        pages = {L14},
          doi = {10.3847/2041-8213/ab0e7a},
archivePrefix = {arXiv},
       eprint = {1902.04689},
 primaryClass = {astro-ph.HE},
       adsurl = {https://ui.adsabs.harvard.edu/abs/2019ApJ...874L..14D}
}

@ARTICLE{Faucher2006,
       author = {{Faucher-Gigu{\`e}re}, Claude-Andr{\'e} and {Kaspi}, Victoria M.},
        title = "{Birth and Evolution of Isolated Radio Pulsars}",
      journal = {\apj},
         year = 2006,
        month = may,
       volume = {643},
       number = {1},
        pages = {332-355},
          doi = {10.1086/501516},
archivePrefix = {arXiv},
       eprint = {astro-ph/0512585},
 primaryClass = {astro-ph},
       adsurl = {https://ui.adsabs.harvard.edu/abs/2006ApJ...643..332F}
}

@ARTICLE{Ruderman1975,
       author = {{Ruderman}, M.~A. and {Sutherland}, P.~G.},
        title = "{Theory of pulsars: polar gaps, sparks, and coherent microwave radiation.}",
      journal = {\apj},
         year = 1975,
        month = feb,
       volume = {196},
        pages = {51-72},
          doi = {10.1086/153393},
       adsurl = {https://ui.adsabs.harvard.edu/abs/1975ApJ...196...51R}
}

@ARTICLE{Kaspi2017,
       author = {{Kaspi}, Victoria M. and {Beloborodov}, Andrei M.},
        title = "{Magnetars}",
      journal = {\araa},
         year = 2017,
        month = aug,
       volume = {55},
       number = {1},
        pages = {261-301},
          doi = {10.1146/annurev-astro-081915-023329},
archivePrefix = {arXiv},
       eprint = {1703.00068},
 primaryClass = {astro-ph.HE},
       adsurl = {https://ui.adsabs.harvard.edu/abs/2017ARA&A..55..261K}
}

@ARTICLE{Lyne1985,
       author = {{Lyne}, A.~G. and {Manchester}, R.~N. and {Taylor}, J.~H.},
        title = "{The galactic population of pulsars.}",
      journal = {\mnras},
         year = 1985,
        month = apr,
       volume = {213},
        pages = {613-639},
          doi = {10.1093/mnras/213.3.613},
       adsurl = {https://ui.adsabs.harvard.edu/abs/1985MNRAS.213..613L}
}

@ARTICLE{Gullon2015,
       author = {{Gull{\'o}n}, M. and {Pons}, J.~A. and {Miralles}, J.~A. and {Vigan{\`o}}, D. and {Rea}, N. and {Perna}, R.},
        title = "{Population synthesis of isolated neutron stars with magneto-rotational evolution - II. From radio-pulsars to magnetars}",
      journal = {\mnras},
         year = 2015,
        month = nov,
       volume = {454},
       number = {1},
        pages = {615-625},
          doi = {10.1093/mnras/stv1644},
archivePrefix = {arXiv},
       eprint = {1507.05452},
 primaryClass = {astro-ph.HE},
       adsurl = {https://ui.adsabs.harvard.edu/abs/2015MNRAS.454..615G}
}

@ARTICLE{Sautron2024,
       author = {{Sautron}, Matt{\'e}o and {P{\'e}tri}, J{\'e}r{\^o}me and {Mitra}, Dipanjan and {Dirson}, Ludmilla},
        title = "{The Galactic population of canonical pulsars: II. Taking into account several evolutionary paths: Interstellar medium interaction, Galactic potential, and a death valley}",
      journal = {\aap},
         year = 2024,
        month = nov,
       volume = {691},
          eid = {A349},
        pages = {A349},
          doi = {10.1051/0004-6361/202451097},
archivePrefix = {arXiv},
       eprint = {2406.12612},
 primaryClass = {astro-ph.HE},
       adsurl = {https://ui.adsabs.harvard.edu/abs/2024A&A...691A.349S}
}

@ARTICLE{Young2010,
       author = {{Young}, M.~D.~T. and {Chan}, L.~S. and {Burman}, R.~R. and {Blair}, D.~G.},
        title = "{Pulsar magnetic alignment and the pulsewidth-age relation}",
      journal = {\mnras},
         year = 2010,
        month = feb,
       volume = {402},
       number = {2},
        pages = {1317-1329},
          doi = {10.1111/j.1365-2966.2009.15972.x},
archivePrefix = {arXiv},
       eprint = {0911.0502},
 primaryClass = {astro-ph.HE},
       adsurl = {https://ui.adsabs.harvard.edu/abs/2010MNRAS.402.1317Y}
}

@ARTICLE{Davis1970,
       author = {{Davis}, Leverett and {Goldstein}, Mark},
        title = "{Magnetic-Dipole Alignment in Pulsars}",
      journal = {\apjl},
         year = 1970,
        month = feb,
       volume = {159},
        pages = {L81-L86},
          doi = {10.1086/180482},
       adsurl = {https://ui.adsabs.harvard.edu/abs/1970ApJ...159L..81D}
}

@ARTICLE{Deutsch1955,
       author = {{Deutsch}, Arnim J.},
        title = "{The electromagnetic field of an idealized star in rigid rotation in vacuo}",
      journal = {Annales d'Astrophysique},
         year = 1955,
        month = jan,
       volume = {18},
        pages = {1},
       adsurl = {https://ui.adsabs.harvard.edu/abs/1955AnAp...18....1D}
}

@ARTICLE{Ostriker1969,
       author = {{Ostriker}, J.~P. and {Gunn}, J.~E.},
        title = "{On the Nature of Pulsars. I. Theory}",
      journal = {\apj},
         year = 1969,
        month = sep,
       volume = {157},
        pages = {1395},
          doi = {10.1086/150160},
       adsurl = {https://ui.adsabs.harvard.edu/abs/1969ApJ...157.1395O}
}

@ARTICLE{Michel1970,
       author = {{Michel}, F.~C. and {Goldwire}, Jr., H.~C.},
        title = "{Alignment of Oblique Rotators}",
      journal = {\aplett},
         year = 1970,
        month = jan,
       volume = {5},
        pages = {21},
       adsurl = {https://ui.adsabs.harvard.edu/abs/1970ApL.....5...21M}
}

@ARTICLE{Li2023,
       author = {{Li}, Biao-Peng and {Gao}, Zhi-Fu},
        title = "{A short review on the pulsar magnetic inclination angles}",
      journal = {Astronomische Nachrichten},
         year = 2023,
        month = aug,
       volume = {344},
       number = {6},
          eid = {e20220111},
        pages = {e20220111},
          doi = {10.1002/asna.20220111},
archivePrefix = {arXiv},
       eprint = {2404.13857},
 primaryClass = {astro-ph.HE},
       adsurl = {https://ui.adsabs.harvard.edu/abs/2023AN....34420111L}
}

@ARTICLE{Beniamini2017,
       author = {{Beniamini}, Paz and {Giannios}, Dimitrios and {Metzger}, Brian D.},
        title = "{Constraints on millisecond magnetars as the engines of prompt emission in gamma-ray bursts}",
      journal = {\mnras},
         year = 2017,
        month = dec,
       volume = {472},
       number = {3},
        pages = {3058-3073},
          doi = {10.1093/mnras/stx2095},
archivePrefix = {arXiv},
       eprint = {1706.05014},
 primaryClass = {astro-ph.HE},
       adsurl = {https://ui.adsabs.harvard.edu/abs/2017MNRAS.472.3058B}
}

@ARTICLE{Li2025,
       author = {{Li}, Biao-Peng and {Gao}, Zhi-Fu and {Ma}, Wen-Qi and {Cheng}, Quan},
        title = "{Braking index of PSR J1846‑0258: a model of magnetic inclination evolution and its gravitational-wave implication}",
      journal = {Frontiers in Astronomy and Space Sciences},
         year = 2025,
        month = jul,
       volume = {12},
          eid = {1625459},
        pages = {1625459},
          doi = {10.3389/fspas.2025.1625459},
       adsurl = {https://ui.adsabs.harvard.edu/abs/2025FrASS..1225459L}
}

@ARTICLE{Cieslar2020,
       author = {{Cie{\'s}lar}, Marek and {Bulik}, Tomasz and {Os{\l}owski}, Stefan},
        title = "{Markov Chain Monte Carlo population synthesis of single radio pulsars in the Galaxy}",
      journal = {\mnras},
         year = 2020,
        month = mar,
       volume = {492},
       number = {3},
        pages = {4043-4057},
          doi = {10.1093/mnras/staa073},
       adsurl = {https://ui.adsabs.harvard.edu/abs/2020MNRAS.492.4043C}
}

@ARTICLE{Sturrock1971,
       author = {{Sturrock}, P.~A.},
        title = "{A Model of Pulsars}",
      journal = {\apj},
         year = 1971,
        month = mar,
       volume = {164},
        pages = {529},
          doi = {10.1086/150865},
       adsurl = {https://ui.adsabs.harvard.edu/abs/1971ApJ...164..529S}
}

@ARTICLE{Manchester2005,
       author = {{Manchester}, R.~N. and {Hobbs}, G.~B. and {Teoh}, A. and {Hobbs}, M.},
        title = "{The Australia Telescope National Facility Pulsar Catalogue}",
      journal = {\aj},
         year = 2005,
        month = apr,
       volume = {129},
       number = {4},
        pages = {1993-2006},
          doi = {10.1086/428488},
archivePrefix = {arXiv},
       eprint = {astro-ph/0412641},
 primaryClass = {astro-ph},
       adsurl = {https://ui.adsabs.harvard.edu/abs/2005AJ....129.1993M}
}

@ARTICLE{Mcmillan2017,
       author = {{McMillan}, Paul J.},
        title = "{The mass distribution and gravitational potential of the Milky Way}",
      journal = {\mnras},
         year = 2017,
        month = feb,
       volume = {465},
       number = {1},
        pages = {76-94},
          doi = {10.1093/mnras/stw2759},
archivePrefix = {arXiv},
       eprint = {1608.00971},
 primaryClass = {astro-ph.GA},
       adsurl = {https://ui.adsabs.harvard.edu/abs/2017MNRAS.465...76M}
}

@ARTICLE{wilm2000,
       author = {{Wilms}, J. and {Allen}, A. and {McCray}, R.},
        title = "{On the Absorption of X-Rays in the Interstellar Medium}",
      journal = {\apj},
         year = 2000,
        month = oct,
       volume = {542},
       number = {2},
        pages = {914-924},
          doi = {10.1086/317016},
archivePrefix = {arXiv},
       eprint = {astro-ph/0008425},
 primaryClass = {astro-ph},
       adsurl = {https://ui.adsabs.harvard.edu/abs/2000ApJ...542..914W}
}

@ARTICLE{Jawor2022,
       author = {{Jawor}, Jedrzej A. and {Tauris}, Thomas M.},
        title = "{Modelling spin evolution of magnetars}",
      journal = {\mnras},
         year = 2022,
        month = jan,
       volume = {509},
       number = {1},
        pages = {634-657},
          doi = {10.1093/mnras/stab2677},
archivePrefix = {arXiv},
       eprint = {2109.07484},
 primaryClass = {astro-ph.HE},
       adsurl = {https://ui.adsabs.harvard.edu/abs/2022MNRAS.509..634J}
}

@ARTICLE{Manchester2001,
       author = {{Manchester}, R.~N. and {Lyne}, A.~G. and {Camilo}, F. and {Bell}, J.~F. and {Kaspi}, V.~M. and {D'Amico}, N. and {McKay}, N.~P.~F. and {Crawford}, F. and {Stairs}, I.~H. and {Possenti}, A. and {Kramer}, M. and {Sheppard}, D.~C.},
        title = "{The Parkes multi-beam pulsar survey - I. Observing and data analysis systems, discovery and timing of 100 pulsars}",
      journal = {\mnras},
         year = 2001,
        month = nov,
       volume = {328},
       number = {1},
        pages = {17-35},
          doi = {10.1046/j.1365-8711.2001.04751.x},
archivePrefix = {arXiv},
       eprint = {astro-ph/0106522},
 primaryClass = {astro-ph},
       adsurl = {https://ui.adsabs.harvard.edu/abs/2001MNRAS.328...17M}
}

@online{Chandra,
    title = {The Chandra Proposers' Observatory Guide},
    author = {Chandra X-ray Center
Chandra Project Science, MSFC
Chandra IPI Teams},
    year = {2024},
    url = {https://cxc.harvard.edu/proposer/POG/html/index.html},
    urldate      = {Use the date of access}
}

@ARTICLE{Perna2011,
       author = {{Perna}, Rosalba and {Pons}, Jose A.},
        title = "{A Unified Model of the Magnetar and Radio Pulsar Bursting Phenomenology}",
      journal = {\apjl},
         year = 2011,
        month = feb,
       volume = {727},
       number = {2},
          eid = {L51},
        pages = {L51},
          doi = {10.1088/2041-8205/727/2/L51},
archivePrefix = {arXiv},
       eprint = {1101.1098},
 primaryClass = {astro-ph.HE},
       adsurl = {https://ui.adsabs.harvard.edu/abs/2011ApJ...727L..51P}
}

@ARTICLE{Igoshev2022,
       author = {{Igoshev}, Andrei P. and {Frantsuzova}, Anastasia and {Gourgouliatos}, Konstantinos N. and {Tsichli}, Savina and {Konstantinou}, Lydia and {Popov}, Sergei B.},
        title = "{Initial periods and magnetic fields of neutron stars}",
      journal = {\mnras},
         year = 2022,
        month = aug,
       volume = {514},
       number = {3},
        pages = {4606-4619},
          doi = {10.1093/mnras/stac1648},
archivePrefix = {arXiv},
       eprint = {2205.06823},
 primaryClass = {astro-ph.HE},
       adsurl = {https://ui.adsabs.harvard.edu/abs/2022MNRAS.514.4606I}
}

@ARTICLE{Enoto2019,
       author = {{Enoto}, Teruaki and {Kisaka}, Shota and {Shibata}, Shinpei},
        title = "{Observational diversity of magnetized neutron stars}",
      journal = {Reports on Progress in Physics},
         year = 2019,
        month = oct,
       volume = {82},
       number = {10},
          eid = {106901},
        pages = {106901},
          doi = {10.1088/1361-6633/ab3def},
       adsurl = {https://ui.adsabs.harvard.edu/abs/2019RPPh...82j6901E}
}

@ARTICLE{Popov2010,
       author = {{Popov}, S.~B. and {Pons}, J.~A. and {Miralles}, J.~A. and {Boldin}, P.~A. and {Posselt}, B.},
        title = "{Population synthesis studies of isolated neutron stars with magnetic field decay}",
      journal = {\mnras},
         year = 2010,
        month = feb,
       volume = {401},
       number = {4},
        pages = {2675-2686},
          doi = {10.1111/j.1365-2966.2009.15850.x},
archivePrefix = {arXiv},
       eprint = {0910.2190},
 primaryClass = {astro-ph.HE},
       adsurl = {https://ui.adsabs.harvard.edu/abs/2010MNRAS.401.2675P}
}

@ARTICLE{Kaspi2005,
       author = {{Kaspi}, V.~M. and {McLaughlin}, M.~A.},
        title = "{Chandra X-Ray Detection of the High Magnetic Field Radio Pulsar PSR J1718-3718}",
      journal = {\apjl},
         year = 2005,
        month = jan,
       volume = {618},
       number = {1},
        pages = {L41-L44},
          doi = {10.1086/427628},
archivePrefix = {arXiv},
       eprint = {astro-ph/0411615},
 primaryClass = {astro-ph},
       adsurl = {https://ui.adsabs.harvard.edu/abs/2005ApJ...618L..41K}
}

@ARTICLE{Espinoza2017,
       author = {{Espinoza}, C.~M. and {Lyne}, A.~G. and {Stappers}, B.~W.},
        title = "{New long-term braking index measurements for glitching pulsars using a glitch-template method}",
      journal = {\mnras},
         year = 2017,
        month = apr,
       volume = {466},
       number = {1},
        pages = {147-162},
          doi = {10.1093/mnras/stw3081},
archivePrefix = {arXiv},
       eprint = {1611.08314},
 primaryClass = {astro-ph.HE},
       adsurl = {https://ui.adsabs.harvard.edu/abs/2017MNRAS.466..147E}
}

@ARTICLE{Suzuki2021,
       author = {{Suzuki}, Hiromasa and {Bamba}, Aya and {Shibata}, Shinpei},
        title = "{Quantitative Age Estimation of Supernova Remanants and Associated Pulsars}",
      journal = {\apj},
         year = 2021,
        month = jun,
       volume = {914},
       number = {2},
          eid = {103},
        pages = {103},
          doi = {10.3847/1538-4357/abfb02},
archivePrefix = {arXiv},
       eprint = {2104.10052},
 primaryClass = {astro-ph.HE},
       adsurl = {https://ui.adsabs.harvard.edu/abs/2021ApJ...914..103S}
}

@ARTICLE{Ferrand2012,
       author = {{Ferrand}, Gilles and {Safi-Harb}, Samar},
        title = "{A census of high-energy observations of Galactic supernova remnants}",
      journal = {Advances in Space Research},
         year = 2012,
        month = may,
       volume = {49},
       number = {9},
        pages = {1313-1319},
          doi = {10.1016/j.asr.2012.02.004},
archivePrefix = {arXiv},
       eprint = {1202.0245},
 primaryClass = {astro-ph.HE},
       adsurl = {https://ui.adsabs.harvard.edu/abs/2012AdSpR..49.1313F}
}

@ARTICLE{Hurley2005,
       author = {{Hurley}, K. and {Boggs}, S.~E. and {Smith}, D.~M. and {Duncan}, R.~C. and {Lin}, R. and {Zoglauer}, A. and {Krucker}, S. and {Hurford}, G. and {Hudson}, H. and {Wigger}, C. and {Hajdas}, W. and {Thompson}, C. and {Mitrofanov}, I. and {Sanin}, A. and {Boynton}, W. and {Fellows}, C. and {von Kienlin}, A. and {Lichti}, G. and {Rau}, A. and {Cline}, T.},
        title = "{An exceptionally bright flare from SGR 1806-20 and the origins of short-duration {\ensuremath{\gamma}}-ray bursts}",
      journal = {\nat},
         year = 2005,
        month = apr,
       volume = {434},
       number = {7037},
        pages = {1098-1103},
          doi = {10.1038/nature03519},
archivePrefix = {arXiv},
       eprint = {astro-ph/0502329},
 primaryClass = {astro-ph},
       adsurl = {https://ui.adsabs.harvard.edu/abs/2005Natur.434.1098H}
}

@ARTICLE{Lower2026,
       author = {{Lower}, Marcus E. and {Scholz}, Paul and {Camilo}, Fernando and {Palmer}, David M. and {Reynolds}, John E. and {Sarkissian}, John M. and {Toomey}, Lawrence J. and {Younes}, George},
        title = "{Transient narrowband radio bursts from 1E 1547.0-5408}",
      journal = {arXiv e-prints},
         year = 2026,
        month = mar,
          eid = {arXiv:2603.21450},
        pages = {arXiv:2603.21450},
          doi = {10.48550/arXiv.2603.21450},
archivePrefix = {arXiv},
       eprint = {2603.21450},
 primaryClass = {astro-ph.HE},
       adsurl = {https://ui.adsabs.harvard.edu/abs/2026arXiv260321450L}
}

@ARTICLE{Cummings2011,
       author = {{Cummings}, J.~R. and {Burrows}, D. and {Campana}, S. and {Kennea}, J.~A. and {Krimm}, H.~A. and {Palmer}, D.~M. and {Sakamoto}, T. and {Zan}, S.},
        title = "{Swift J1822.3-1606: A Probable New SGR in Ground Analysis of BAT Data}",
      journal = {The Astronomer's Telegram},
         year = 2011,
        month = jul,
       volume = {3488},
        pages = {1},
       adsurl = {https://ui.adsabs.harvard.edu/abs/2011ATel.3488....1C}
}

@ARTICLE{Horst2010,
       author = {{van der Horst}, A.~J. and {Connaughton}, V. and {Kouveliotou}, C. and {G{\"o}{\v{g}}{\"u}{\textcommabelow s}}, E. and {Kaneko}, Y. and {Wachter}, S. and {Briggs}, M.~S. and {Granot}, J. and {Ramirez-Ruiz}, E. and {Woods}, P.~M. and {Aptekar}, R.~L. and {Barthelmy}, S.~D. and {Cummings}, J.~R. and {Finger}, M.~H. and {Frederiks}, D.~D. and {Gehrels}, N. and {Gelino}, C.~R. and {Gelino}, D.~M. and {Golenetskii}, S. and {Hurley}, K. and {Krimm}, H.~A. and {Mazets}, E.~P. and {McEnery}, J.~E. and {Meegan}, C.~A. and {Oleynik}, P.~P. and {Palmer}, D.~M. and {Pal'shin}, V.~D. and {Pe'er}, A. and {Svinkin}, D. and {Ulanov}, M.~V. and {van der Klis}, M. and {von Kienlin}, A. and {Watts}, A.~L. and {Wilson-Hodge}, C.~A.},
        title = "{Discovery of a New Soft Gamma Repeater: SGR J0418 + 5729}",
      journal = {\apjl},
         year = 2010,
        month = mar,
       volume = {711},
       number = {1},
        pages = {L1-L6},
          doi = {10.1088/2041-8205/711/1/L1},
archivePrefix = {arXiv},
       eprint = {0911.5544},
 primaryClass = {astro-ph.HE},
       adsurl = {https://ui.adsabs.harvard.edu/abs/2010ApJ...711L...1V}
}

@ARTICLE{Gogus2000,
   author = {{G{\"o}{\v g}{\"u}{\c s}}, E. and {Woods}, P.~M. and {Kouveliotou}, C. and 
	{van Paradijs}, J. and {Briggs}, M.~S. and {Duncan}, R.~C. and 
	{Thompson}, C.},
    title = "{Statistical Properties of SGR 1806-20 Bursts}",
  journal = {\apjl},
   eprint = {astro-ph/0002181},
     year = 2000,
    month = apr,
   volume = 532,
    pages = {L121-L124},
      doi = {10.1086/312583},
   adsurl = {http://adsabs.harvard.edu/abs/2000ApJ...532L.121G}
}

@ARTICLE{Lien2016,
       author = {{Lien}, Amy and {Sakamoto}, Takanori and {Barthelmy}, Scott D. and {Baumgartner}, Wayne H. and {Cannizzo}, John K. and {Chen}, Kevin and {Collins}, Nicholas R. and {Cummings}, Jay R. and {Gehrels}, Neil and {Krimm}, Hans A. and {Markwardt}, Craig. B. and {Palmer}, David M. and {Stamatikos}, Michael and {Troja}, Eleonora and {Ukwatta}, T.~N.},
        title = "{The Third Swift Burst Alert Telescope Gamma-Ray Burst Catalog}",
      journal = {\apj},
         year = 2016,
        month = sep,
       volume = {829},
       number = {1},
          eid = {7},
        pages = {7},
          doi = {10.3847/0004-637X/829/1/7},
archivePrefix = {arXiv},
       eprint = {1606.01956},
 primaryClass = {astro-ph.HE},
       adsurl = {https://ui.adsabs.harvard.edu/abs/2016ApJ...829....7L}
}

@ARTICLE{Kienlin2020,
       author = {{von Kienlin}, A. and {Meegan}, C.~A. and {Paciesas}, W.~S. and {Bhat}, P.~N. and {Bissaldi}, E. and {Briggs}, M.~S. and {Burns}, E. and {Cleveland}, W.~H. and {Gibby}, M.~H. and {Giles}, M.~M. and {Goldstein}, A. and {Hamburg}, R. and {Hui}, C.~M. and {Kocevski}, D. and {Mailyan}, B. and {Malacaria}, C. and {Poolakkil}, S. and {Preece}, R.~D. and {Roberts}, O.~J. and {Veres}, P. and {Wilson-Hodge}, C.~A.},
        title = "{The Fourth Fermi-GBM Gamma-Ray Burst Catalog: A Decade of Data}",
      journal = {\apj},
         year = 2020,
        month = apr,
       volume = {893},
       number = {1},
          eid = {46},
        pages = {46},
          doi = {10.3847/1538-4357/ab7a18},
archivePrefix = {arXiv},
       eprint = {2002.11460},
 primaryClass = {astro-ph.HE},
       adsurl = {https://ui.adsabs.harvard.edu/abs/2020ApJ...893...46V}
}

@ARTICLE{Feroci2001,
       author = {{Feroci}, M. and {Hurley}, K. and {Duncan}, R.~C. and {Thompson}, C.},
        title = "{The Giant Flare of 1998 August 27 from SGR 1900+14. I. An Interpretive Study of BeppoSAX and Ulysses Observations}",
      journal = {\apj},
         year = 2001,
        month = mar,
       volume = {549},
       number = {2},
        pages = {1021-1038},
          doi = {10.1086/319441},
archivePrefix = {arXiv},
       eprint = {astro-ph/0010494},
 primaryClass = {astro-ph},
       adsurl = {https://ui.adsabs.harvard.edu/abs/2001ApJ...549.1021F}
}

@ARTICLE{Mazets1979,
       author = {{Mazets}, E.~P. and {Golentskii}, S.~V. and {Ilinskii}, V.~N. and {Aptekar}, R.~L. and {Guryan}, Iu. A.},
        title = "{Observations of a flaring X-ray pulsar in Dorado}",
      journal = {\nat},
         year = 1979,
        month = dec,
       volume = {282},
       number = {5739},
        pages = {587-589},
          doi = {10.1038/282587a0},
       adsurl = {https://ui.adsabs.harvard.edu/abs/1979Natur.282..587M}
}

@ARTICLE{Kaplan2001,
       author = {{Kaplan}, D.~L. and {Kulkarni}, S.~R. and {van Kerkwijk}, M.~H. and {Rothschild}, R.~E. and {Lingenfelter}, R.~L. and {Marsden}, D. and {Danner}, R. and {Murakami}, T.},
        title = "{Hubble Space Telescope Observations of SGR 0526-66: New Constraints on Accretion and Magnetar Models}",
      journal = {\apj},
         year = 2001,
        month = jul,
       volume = {556},
       number = {1},
        pages = {399-407},
          doi = {10.1086/323516},
archivePrefix = {arXiv},
       eprint = {astro-ph/0103179},
 primaryClass = {astro-ph},
       adsurl = {https://ui.adsabs.harvard.edu/abs/2001ApJ...556..399K}
}

@ARTICLE{Palmer2005,
       author = {{Palmer}, D.~M. and {Barthelmy}, S. and {Gehrels}, N. and {Kippen}, R.~M. and {Cayton}, T. and {Kouveliotou}, C. and {Eichler}, D. and {Wijers}, R.~A.~M.~J. and {Woods}, P.~M. and {Granot}, J. and {Lyubarsky}, Y.~E. and {Ramirez-Ruiz}, E. and {Barbier}, L. and {Chester}, M. and {Cummings}, J. and {Fenimore}, E.~E. and {Finger}, M.~H. and {Gaensler}, B.~M. and {Hullinger}, D. and {Krimm}, H. and {Markwardt}, C.~B. and {Nousek}, J.~A. and {Parsons}, A. and {Patel}, S. and {Sakamoto}, T. and {Sato}, G. and {Suzuki}, M. and {Tueller}, J.},
        title = "{A giant {\ensuremath{\gamma}}-ray flare from the magnetar SGR 1806 - 20}",
      journal = {\nat},
         year = 2005,
        month = apr,
       volume = {434},
       number = {7037},
        pages = {1107-1109},
          doi = {10.1038/nature03525},
archivePrefix = {arXiv},
       eprint = {astro-ph/0503030},
 primaryClass = {astro-ph},
       adsurl = {https://ui.adsabs.harvard.edu/abs/2005Natur.434.1107P}
}

@ARTICLE{Jansen2001,
       author = {{Jansen}, F. and {Lumb}, D. and {Altieri}, B. and {Clavel}, J. and {Ehle}, M. and {Erd}, C. and {Gabriel}, C. and {Guainazzi}, M. and {Gondoin}, P. and {Much}, R. and {Munoz}, R. and {Santos}, M. and {Schartel}, N. and {Texier}, D. and {Vacanti}, G.},
        title = "{XMM-Newton observatory. I. The spacecraft and operations}",
      journal = {\aap},
         year = 2001,
        month = jan,
       volume = {365},
        pages = {L1-L6},
          doi = {10.1051/0004-6361:20000036},
       adsurl = {https://ui.adsabs.harvard.edu/abs/2001A&A...365L...1J}
}

@INPROCEEDINGS{deluca2017,
       author = {{De Luca}, A.},
        title = "{Central compact objects in supernova remnants}",
    booktitle = {Journal of Physics Conference Series},
         year = 2017,
       series = {Journal of Physics Conference Series},
       volume = {932},
        month = dec,
    publisher = {IOP},
          eid = {012006},
        pages = {012006},
          doi = {10.1088/1742-6596/932/1/012006},
archivePrefix = {arXiv},
       eprint = {1711.07210},
 primaryClass = {astro-ph.HE},
       adsurl = {https://ui.adsabs.harvard.edu/abs/2017JPhCS.932a2006D}
}

@ARTICLE{Vigano2012,
       author = {{Vigan{\`o}}, D. and {Pons}, J.~A.},
        title = "{Central compact objects and the hidden magnetic field scenario}",
      journal = {\mnras},
         year = 2012,
        month = oct,
       volume = {425},
       number = {4},
        pages = {2487-2492},
          doi = {10.1111/j.1365-2966.2012.21679.x},
archivePrefix = {arXiv},
       eprint = {1206.2014},
 primaryClass = {astro-ph.SR},
       adsurl = {https://ui.adsabs.harvard.edu/abs/2012MNRAS.425.2487V}
}

@ARTICLE{Shabaltas2012,
       author = {{Shabaltas}, Natalia and {Lai}, Dong},
        title = "{The Hidden Magnetic Field of the Young Neutron Star in Kesteven 79}",
      journal = {\apj},
         year = 2012,
        month = apr,
       volume = {748},
       number = {2},
          eid = {148},
        pages = {148},
          doi = {10.1088/0004-637X/748/2/148},
archivePrefix = {arXiv},
       eprint = {1110.3129},
 primaryClass = {astro-ph.HE},
       adsurl = {https://ui.adsabs.harvard.edu/abs/2012ApJ...748..148S}
}

@ARTICLE{Kennea2013,
       author = {{Kennea}, J.~A. and {Burrows}, D.~N. and {Kouveliotou}, C. and {Palmer}, D.~M. and {G{\"o}{\u{g}}{\"u}{\textcommabelow s}}, E. and {Kaneko}, Y. and {Evans}, P.~A. and {Degenaar}, N. and {Reynolds}, M.~T. and {Miller}, J.~M. and {Wijnands}, R. and {Mori}, K. and {Gehrels}, N.},
        title = "{Swift Discovery of a New Soft Gamma Repeater, SGR J1745-29, near Sagittarius A*}",
      journal = {\apjl},
         year = 2013,
        month = jun,
       volume = {770},
       number = {2},
          eid = {L24},
        pages = {L24},
          doi = {10.1088/2041-8205/770/2/L24},
archivePrefix = {arXiv},
       eprint = {1305.2128},
 primaryClass = {astro-ph.HE},
       adsurl = {https://ui.adsabs.harvard.edu/abs/2013ApJ...770L..24K}
}

@ARTICLE{Tian2007,
       author = {{Tian}, W.~W. and {Li}, Z. and {Leahy}, D.~A. and {Wang}, Q.~D.},
        title = "{VLA and XMM-Newton Observations of the SNR W41/TeV Gamma-Ray Source HESS J1834-087}",
      journal = {\apjl},
         year = 2007,
        month = mar,
       volume = {657},
       number = {1},
        pages = {L25-L28},
          doi = {10.1086/512544},
archivePrefix = {arXiv},
       eprint = {astro-ph/0612296},
 primaryClass = {astro-ph},
       adsurl = {https://ui.adsabs.harvard.edu/abs/2007ApJ...657L..25T}
}

@ARTICLE{nanda2012,
       author = {{Rea}, N. and {Israel}, G.~L. and {Esposito}, P. and {Pons}, J.~A. and {Camero-Arranz}, A. and {Mignani}, R.~P. and {Turolla}, R. and {Zane}, S. and {Burgay}, M. and {Possenti}, A. and {Campana}, S. and {Enoto}, T. and {Gehrels}, N. and {G{\"o}{\v{g}}{\"u}{\textcommabelow s}}, E. and {G{\"o}tz}, D. and {Kouveliotou}, C. and {Makishima}, K. and {Mereghetti}, S. and {Oates}, S.~R. and {Palmer}, D.~M. and {Perna}, R. and {Stella}, L. and {Tiengo}, A.},
        title = "{A New Low Magnetic Field Magnetar: The 2011 Outburst of Swift J1822.3-1606}",
      journal = {\apj},
         year = 2012,
        month = jul,
       volume = {754},
       number = {1},
          eid = {27},
        pages = {27},
          doi = {10.1088/0004-637X/754/1/27},
archivePrefix = {arXiv},
       eprint = {1203.6449},
 primaryClass = {astro-ph.GA},
       adsurl = {https://ui.adsabs.harvard.edu/abs/2012ApJ...754...27R}
}

\appendix




\section{Luminosity correction}\label{sec:area-weighted}
\label{sec:correction}

To reconstruct the intrinsic neutron star population, we evaluate the observable volume of an observed neutron star $i$ with a given observed luminosity $L_i$. We introduce the maximum observable distance $d_{i,\max}$ for each object, defined as the distance out to which $L_i$ would be detectable under a given survey flux sensitivity, $F_{\nu,\lim}$. This provides a quantitative mapping from luminosity to an effective observable volume, allowing a direct assessment of whether frequency-dependent selection effects alone can reproduce the observed gap.

Using $d_{i,\max}$, the expected number of neutron stars in the Galaxy that could exist with luminosity $L_i$ is written as
\begin{align}
    \label{eq:w_i}w_i =
    \frac{\displaystyle \int \rho(R,\phi,z)\, dV}
         {\displaystyle \int_{V(d_{i,\max})} \rho(R,\phi,z)\, dV},
\end{align}
where $\rho(R,\phi,z)$ is the Galactic neutron star number density distribution in spherical coordinates from the galactic centre with the assumption that neutron stars are born along the galactic spiral arms \citep{Faucher2006,Cieslar2020}.

For radio, the maximum observable distance is estimated by
$d_{i,\max}=\sqrt{L_{\mathrm{radio},\,i}/F_{\mathrm{radio,\,lim}}}$. 
For X-rays, the observable distance is estimated by $d_{i,\max}=\sqrt{L_{\mathrm{X},i}\, e^{-\tau_i}/4\pi F_{\mathrm{X,\,lim}}}$,
where $\tau_i$ is the optical depth of interstellar absorption,
$\tau_i=\sigma_{\mathrm{2\,keV}}
    \int_{0}^{d_i}
    n_{\mathrm{H}}(l)\, dl$, 
with $n_{\mathrm{H}}$ being the interstellar number density along the line-of-sight $l$ based on the model of a thin disc \citep{Mcmillan2017}, and $d_i$ the observed distance of the neutron star. $\sigma_{2\,\mathrm{keV}}$ is the cross section at $2\,\mathrm{keV}$ \citep{wilm2000}. Here we adopt $F_{\mathrm{radio,\,lim}} = 0.1\,\mathrm{mJy}$ and $F_{\mathrm{X,\,lim}} = 10^{-14}\,\mathrm{erg/cm^2/s}$ based on Parkes Multibeam Pulsar Survey \citep{Manchester2001} and {\it Chandra} or {\it XMM-Newton}  flux sensitivity \citep{Hurley2005,Jansen2001}.



 Figure \ref{fig:B_hist_are_weighted_beaming} ({\it{left}}) shows the magnetic field distribution of the observed neutron stars counted with weights based on Eq. (\ref{eq:w_i}). Even after accounting only for radio and X-ray detectability, the observed gap is not alleviated; instead, as seen in the figure, the gap becomes even more pronounced. 


\section{Examples of magnetic inclination angle and magnetic field evolution}

\begin{figure}
    \centering
    \includegraphics[width=0.95\linewidth]{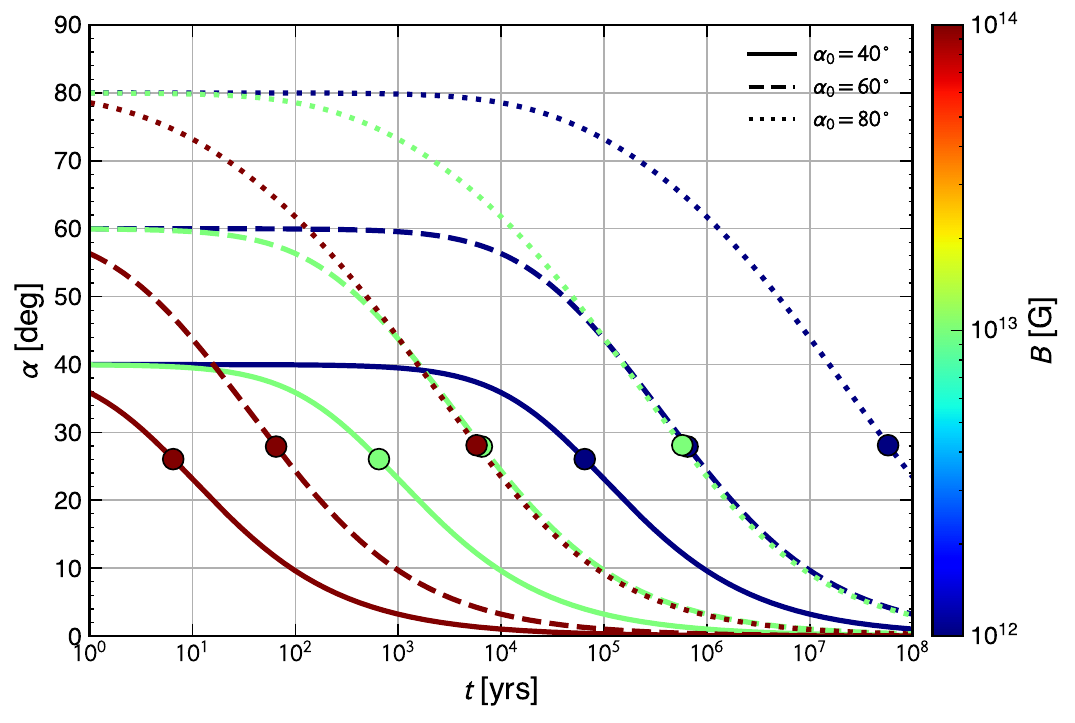}
    \caption{Evolution of the magnetic inclination angle for different initial parameters $B$ and $\alpha_0$. Here we adopt $P_0 = 0.05\,\mathrm{s}$. The point shows the alignment timescale  estimated by Eq. (\ref{eq:tau_alpha}). }
    \label{fig:angle}
\end{figure}

\begin{figure}
    \centering
    \includegraphics[width=0.95\linewidth]{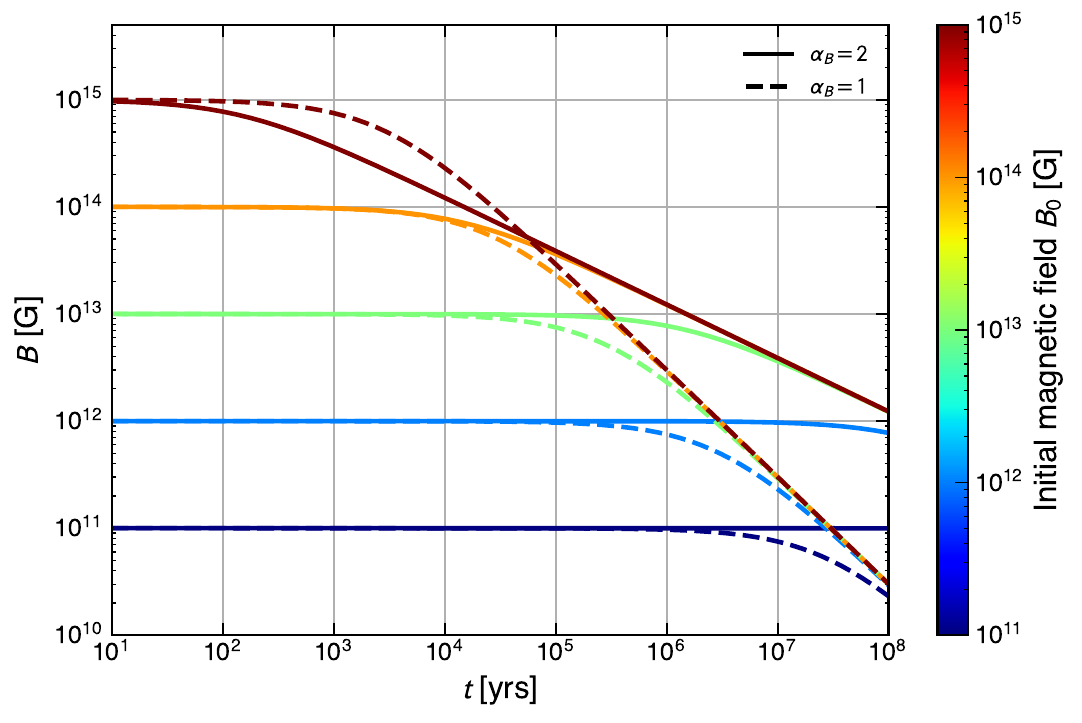}
    \caption{Evolution of the magnetic field with $\alpha_B = 2$ (solid line) and $\alpha_B = 1$ (dotted line). We adopt $\tau_{B,14} = 30\,\mathrm{kyr}$. }
    \label{fig:B}
\end{figure}

\begin{figure}
    \centering
    \includegraphics[width=0.9\linewidth]{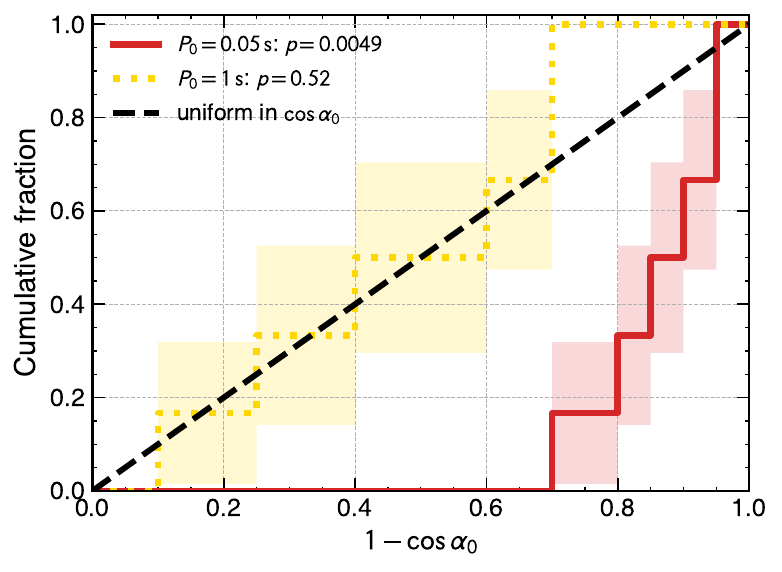}
    \caption{Reconstructed $\alpha_0$ distributions of magnetar for $P_{0}=0.05\,{\rm s}$ (red solid line) and $1\,{\rm s}$ (yellow dotted line) with $\alpha_B = 2$. The dashed line shows an isotropic distribution corresponding to uniform in $\cos\alpha_0$. The KS-test $p$-values are $p=0.0049$ for $P_{0}=0.05\,{\rm s}$ and $p=0.52$ for $P_{0}=1\,{\rm s}$. 
    }
    \label{fig:alpha0_distribution}
\end{figure}

\begin{figure}
    \centering
    \includegraphics[width=0.9\linewidth]{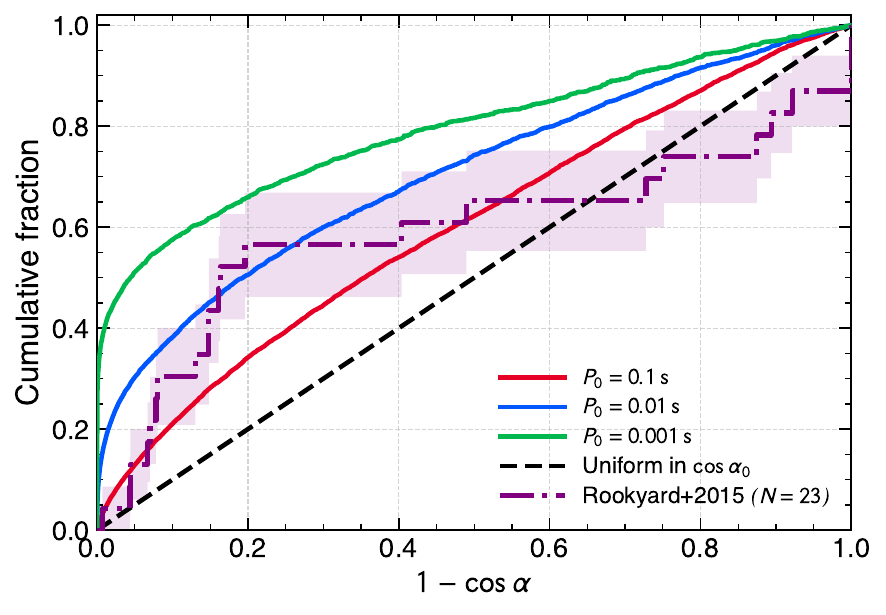}
    \caption{Simulated magnetic inclination angle distributions corrected by beaming fraction for fixed $P_0$, assuming $\log (B_0/\mathrm{G})\sim\mathcal{U}({12},{14})$, $\alpha_B=1$, $\tau_{B,14}=30\,\mathrm{kyr}$, and $t\sim\mathcal{U}(1,10^6)\,\mathrm{yr}$. The observed magnetic inclination angle distributions are also plotted based on \citet{Rookyard2015a}.
    }
    \label{fig:alpha_distribution}
\end{figure}

Figure \ref{fig:angle} shows examples of the time evolution of the magnetic inclination angle for several sets of $B$ and $\alpha_0$ based on Eq. (\ref{eq:Pdot}) and (\ref{eq:alphadot}).
In addition, Figure~\ref{fig:B} shows the magnetic field evolution based on Eq. (\ref{eq:B}) under the assumption that $\tau_B$ depends on the magnetic field. Figure~\ref{fig:alpha0_distribution} shows the reconstructed cumulative initial magnetic inclination angle distributions for $P_{0}=0.05\,{\rm s}$ and $1\,{\rm s}$, together with the theoretical expectation for an isotropic distribution. Figure~\ref{fig:alpha_distribution} shows the cumulative magnetic inclination-angle distributions weighted by the beaming fraction for different initial spin periods. Since $\tau_\alpha \propto P_0^2$, as shown in Eq.~(\ref{eq:tau_alpha}), neutron stars born with shorter spin periods undergo more rapid alignment and therefore exhibit smaller inclination angles.

\section{Luminosity and beaming corrections for $10^4\,\mathrm{yr} < \min (\tau_P, \tau_B) \leq 10^5\,\mathrm{yr}$}

\begin{figure*}
    \centering
    \includegraphics[width = 0.9\linewidth]{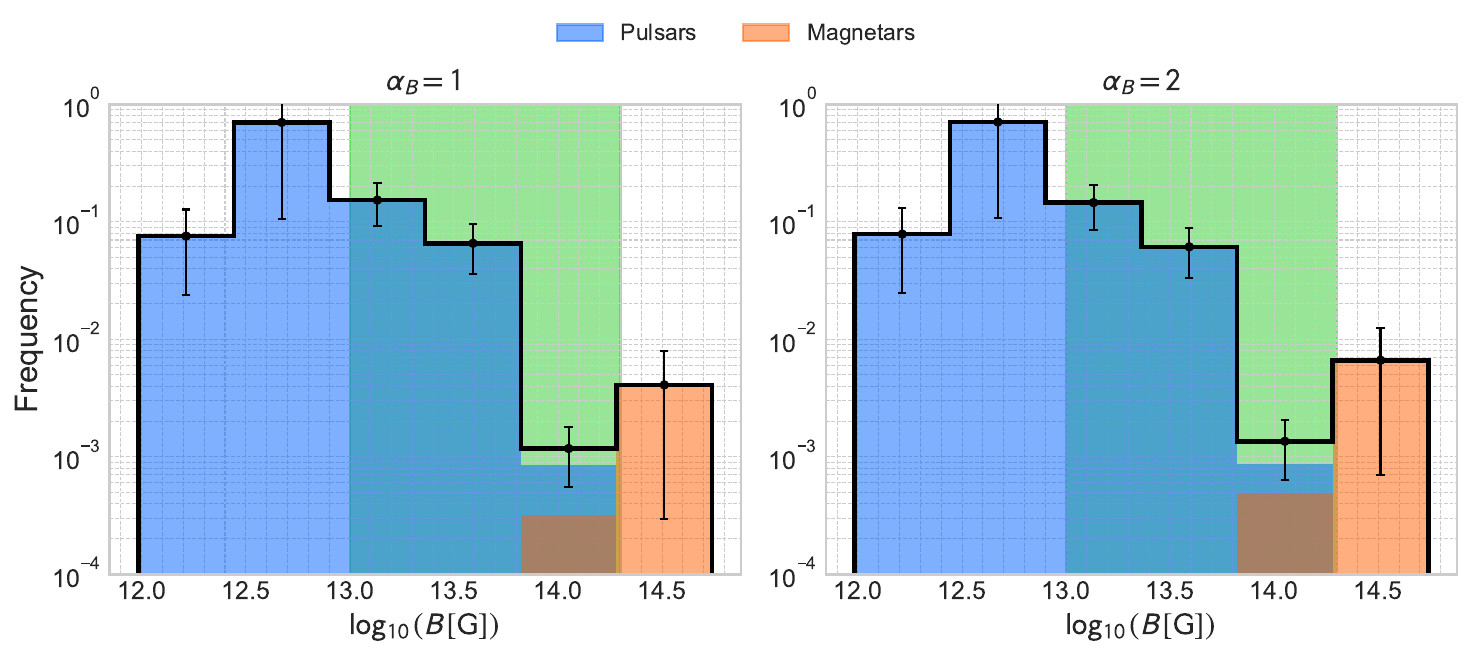}
    \caption{Same as Figure \ref{fig:B_hist_are_weighted_beaming} ({\it{middle}}), but restricted $10^4<\min(\tau_P,\tau_{B,\,\mathrm{p}})\leq10^5\ \mathrm{yr}$.}
    \label{fig:mag_dis1e41e5}
\end{figure*}

As a sanity check, we present Figure~\ref{fig:mag_dis1e41e5}, which shows the magnetic field distribution for the sample restricted to $10^4\,{\rm yr}<\min(\tau_P,\tau_{B,\,{\rm p}})\leq 10^5\,{\rm yr}$. The fraction of magnetars is lower than that of pulsars by roughly two order of magnitude. This is likely because magnetic field decay operates efficiently in the magnetar region over this age interval, reducing the number of observable neutron stars. In contrast, pulsars have longer initial magnetic-field decay timescales, so their magnetic fields have not yet decayed significantly and they continue to populate the $P$--$\dot P$ diagram.

\section{Analysis based on SNR age}

\begin{figure*}
    \centering
    \includegraphics[width=0.9\linewidth]{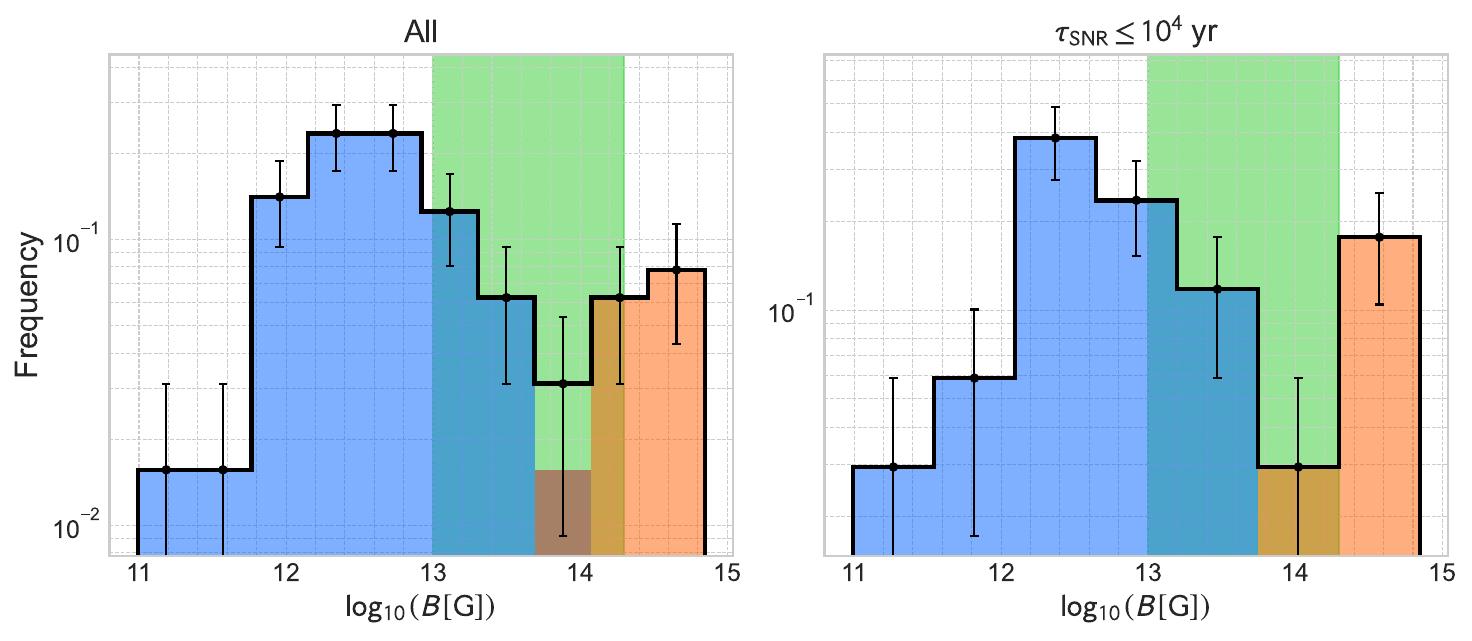}
    \caption{Same as Figure \ref{fig:B_hist}, but for SNR associated neutron stars.}
    \label{fig:B_hist_snr}
\end{figure*}

\begin{figure*}
    \centering\includegraphics[width=0.95\linewidth]{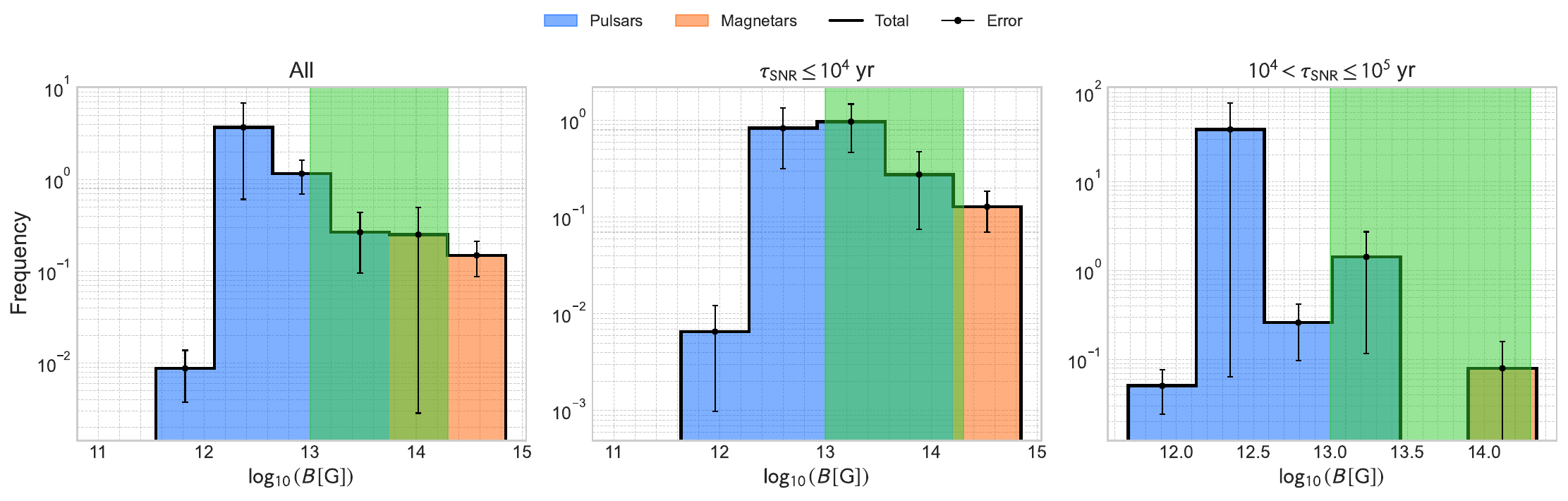}
    \caption{Same as Figure \ref{fig:B_hist_are_weighted_beaming} ({\it{middle}}), but for SNR associated neutron stars.}
    \label{fig:B_hist_beaming_snr}
\end{figure*}

Observationally, the braking index of neutron stars is smaller than the dipole radiation, $n=3$, and in some cases is as low as $n \sim 1$ \citep{Espinoza2017}. This implies that the spin-down age does not necessarily provide the true age. Therefore, analyses based on the spin-down age may be affected by systematic uncertainties. Based on that, it is crucial to reanalyze using SNR ages, which are expected to be closer to the true ages \citep{Suzuki2021}. We restrict the sample to pulsars and magnetars associated with SNR based on SNRcat \citep{Ferrand2012} and correct with the same method as we see in Appendix \ref{sec:correction} and \ref{sec:result} in Figure \ref{fig:B_hist_snr} and \ref{fig:B_hist_beaming_snr}. There is no significant difference in either the observed magnetic field distribution between the SNR age and minimum of the spin-down age and the magnetic decay timescale. This suggests that systematic uncertainties in age estimates have only a minor effect on our interpretation of the gap in the alignment scenario.


\bsp	
\label{lastpage}
\end{document}